\documentclass[preprint2]{aastex}

\usepackage{hyperref}

\usepackage{mathtools}
\usepackage{natbib}
\usepackage{graphicx, subfigure}
\usepackage{multirow}
\usepackage{color}
\usepackage{amsmath}
\usepackage{ulem}
\newcommand{\angstrom}{\textup{\AA}}
\usepackage{verbatim}
\usepackage[colorinlistoftodos]{todonotes}

\shorttitle{RV content of Barnard's star spectrum}
\shortauthors{Artigau et al.}

\begin{document}

\title{ Optical and Near-Infrared Radial Velocity Content of M Dwarfs: Testing Models with Barnard's Star.}

\author{ {\'E}tienne Artigau\altaffilmark{1}, Lison Malo\altaffilmark{1,2}, Ren\'e Doyon\altaffilmark{1}, Pedro Figueira\altaffilmark{3,4}, Xavier Delfosse\altaffilmark{5}, Nicola Astudillo-Defru\altaffilmark{6,7} }

\altaffiltext{1}{Institut de Recherche sur les Exoplan{\`e}tes (IREx), D\'epartement de Physique, Universit\'e de Montr\'eal, C.P. 6128, Succ. Centre-Ville, Montr\'eal, QC, H3C 3J7, Canada}
\altaffiltext{2}{Canada-france-Hawaii Corporation, 65-1238 Mamalahoa Hwy, Kamuela, HI, USA, 96743}
\altaffiltext{3}{European Southern Observatory, Alonso de Cordova 3107, Vitacura, Santiago, Chile}
\altaffiltext{4}{Instituto de Astrof\'isica e Ci\^encias do Espa\c{c}o, Universidade do Porto, CAUP, Rua das Estrelas, 4150-762 Porto, Portugal}
\altaffiltext{5}{Universit\'e Grenoble Alpes, CNRS, IPAG, 38000 Grenoble, France}
\altaffiltext{6}{Universidad de Concepci\'on, Departamento de Astronom\'ia, Casilla 160-C, Concepci\'on, Chile}
\altaffiltext{7}{Observatoire de Gen\`eve, Universit\'e de Gen\`eve, 51 ch. des Maillettes, 1290 Sauverny, Switzerland}


\email{Send correspondence to artigau@astro.umontreal.ca}

\begin{abstract}

High precision radial velocity (RV) measurements have been central in the study of exoplanets during the last two decades, from the early discovery of hot Jupiters, to the recent mass measurements of Earth-sized planets uncovered by transit surveys. While optical radial-velocity is now a mature field, there is currently a strong effort to push the technique into the near-infrared (nIR) domain (chiefly $Y$, $J$, $H$ and $K$ band passes) to probe planetary systems around late-type stars. The combined lower mass and luminosity of M dwarfs leads to an increased reflex RV signal for planets in the habitable zone compared to Sun-like stars. The estimates on the detectability of planets rely on various instrumental characteristics, but also on a prior knowledge of the stellar spectrum. While the overall properties of M dwarf spectra have been extensively tested against observations, the same is not true for their detailed line profiles, which leads to significant uncertainties when converting a given signal-to-noise ratio to a corresponding RV precision as attainable on a given spectrograph. By combining archival CRIRES and HARPS data with ESPaDOnS data of Barnard's star, we show that state-of-the-art atmosphere models over-predict the $Y$ and $J$-band RV content by more than a factor of $\sim$$2$, while under-predicting the $H$ and $K$-band content by half.
\end{abstract}

\keywords{techniques: radial velocities, instrumentation: spectrographs, methods: data analysis, stars: low-mass}

\section{Introduction and context}

Radial velocity (RV) measurements at an ever increasing precision have been central to our quest to find planets around other stars. The first exoplanet around a Sun-like star, 51 Pegasi, was found through precise monitoring of its parent star velocity \citep{Mayor:1995} using the ELODIE spectrograph. Over the first decade following the discovery of 51 Pegasi, radial velocity monitoring accounted for 88\%\footnote{\href{http://exoplanets.org}{http://exoplanets.org}} of exoplanet discoveries. With the launch of Corot and Kepler, and thanks to a number of ground-based surveys, the bulk of exoplanet discoveries now come from transit detection. The upcoming launch of the Transiting Exoplanet Survey Satellite (TESS; \citealt{Ricker:2014}) will provide an even larger sample of short-period transiting planets around relatively bright stars over most of the sky. Ground-based RV follow-up of TESS discoveries will require a significant investment of observing time. RV searches continue to play a crucial role in the field by discovering the majority of the planets in the close solar neighborhood, which are not transiting, and in identifying long-period planets, to which transit searches are not sensitive due to the decreasing likelihood of transits on wide orbits and sparseness of transits in time. RV is the prime tool to establish planetary system architectures out to a few astronomical units.   In most cases RV monitoring is the only tool currently available to confirm transiting planets, measure their masses, and ultimately constrain their bulk density. Furthermore, RV measurements can lead to the discovery of additional non-transiting planets in systems with a transiting companion \citep{Cloutier:2017}.

Only a handful of transiting Earth-sized planets orbiting Sun-like stars are known due to the rarity of their occurrence (once a year) and shallow depth. No such planet has been found by RV surveys and none of the transiting ones can be followed-up with current RV instrumentation due to the inherent high-precision required (e.g., the Earth produces an RV reflex motion on the Sun with a semi-amplitude of only 10\,cm/s). Characterization of Earth-like planets is one of the major goals of exoplanet science in the near-future, and M dwarfs represent a short-cut for finding such planets with existing technologies. The interest of M dwarf in the quest of habitable worlds is many-fold. Firstly, despite having a very incomplete census of nearby M-dwarf planets, transit surveys provide strong constraints on the occurrence rate of planets around early M dwarfs. Within the Kepler dataset, \citet{Dressing:2015} derived an occurrence rate for  $1-4$\,R$_\oplus$ planets of 2.5 per star, including an average of 0.56 Earth-sized planet ($1-1.5$\,R$_{\oplus}$) within a 50\,day orbit around Kepler M dwarfs . This result constrasts with the absence of Jupiter-mass planets in radial-velocity searches around early-M dwarfs (occurence $<1$\% for $10^3-10^4$\,M$_{\oplus}$; \citealt{Bonfils:2013}). Neptune-sized and smaller planets appear to abound, implying that the bulk of the nearest exoplanets orbit M dwarfs. This abundance is exemplified by the discovery of an Earth-mass planet in the temperate zone of our closest stellar neighbor: Proxima Centauri \citep{Anglada-Escude:2016}. Besides being ubiquitous, M dwarf planets also have a number of observational advantages easing their study compared to planets orbiting Sun-like stars. The smaller radius of M dwarfs ($0.1-0.5$R$_\odot$) leads to much deeper transit depths for a given planetary radius ; this not only makes the discovery of transiting planets more likely, but also facilitates transit spectroscopy. Planets around M dwarfs in the solar neighborhood will be also the first priority target to characterize atmosphere in coupling high contrasts imaging and high spectral resolution capacity of ELT \citep{Snellen:2015} or even 10-m telescopes \citep{Lovis:2016}. The smaller radius and lower temperature (2500-3900\,K; \citealt{Rajpurohit:2013}) leads to orbital separation for habitable zone planets in the range of 0.017 to 0.2\,AU, which increases the likelihood of transit and leads to short orbital periods ($<100$\,days) for these type of planets. Importantly, the tighter orbit and lower-mass primary (0.07-0.5 M$_\odot$) lead to a much larger radial-velocity signal than for a planet around a Sun-like stars. An Earth-mass planet around a field M5V star ($\sim$0.15R$_\odot$, $\sim0.1$M$_\odot$, 3200\,K) receiving the same illumination as the Earth has an orbital separation of 0.04\,AU and an orbital period of 9 days and induces a radial-velocity signal of 1\,m/s. Signals at this level can be detected by state-of-the-art velocimeters such as HARPS \citep{Pepe:2000a} or Keck HIRES \citep{Vogt:1994}. The main drawback facing existing precision radial velocity (PRV) spectrograph arises from the faintness of M dwarfs at optical wavelengths. The HARPS M dwarf planet search \citep{Bonfils:2013} yielded $\sim$m/s RV precision in 15\,min for stars brighter than $V=10$. Despite the fact that M dwarfs largely outnumber Sun-like stars in the solar neighborhood, there are only $116$ M dwarfs this bright, mostly of early ($<$M4) spectral types (See Section~4.4 in \citealt{Figueira:2016} for a discussion of the local M dwarf sample).

The best strategy to study a large sample of M dwarf HZ planets is to obtain radial velocity measurements with m/s-level, or better, precision in the wavelength domains where the spectral energy distribution of these objects peaks, namely the  near-infrared (nIR; $1-2.4\mu$m). A number  of high-precision RV spectrographs will be commissioned in the near future (e.g., SPIRou, \citet{Artigau:2014d}; IRD, \citet{Tamura:2012}; HPF, \citet{Mahadevan:2014}; NIRPS, \citet{Bouchy:2017}, or have recently seen first light at the telescope (CARMENES-IR, \citet{Quirrenbach:2010}). Other instrument designs favor a far-red design ($0.7-1.0\mu$m), which is competitive with nIR instruments through most of the M dwarf regime except for very late-M dwarfs (e.g., Maroon-X, PARAS, CARMENES-Optical; \citet{Seifahrt:2016, Chakraborty:2014}). The expected sensitivity of a radial-velocity spectrograph depends on its ability to detect the induced RV Doppler effects at a level orders of magnitudes smaller than the instrument's resolution or the natural width of stellar features.

 In terms of instrumental development, there are numerous stability and calibration challenges (see the review by \citealt{Fischer:2016} and references therein), but with intrinsically stable and well-characterized instruments, this floor in the performances can be significantly smaller than astrophysical RV jitter and limitation due to the finite signal-to-noise ratio (SNR) of observations.  In the ideal case of a slightly active, slowly rotating star, the limitation is determined by the radial velocity information content of the stellar spectra and the SNR with which is achieved, and the purpose of the present paper is to assess this limitation for various instrument setups. \citet{Connes:1985} and \cite{Bouchy:2001} provide a formal framework for determining the ultimate radial velocity precision that can be reached for a given spectrum and SNR.

Modeled M dwarf spectra can be used to derive expected sensitivities for existing and upcoming nIR precision radial velocity (pRV) spectrographs (e.g., \citealt{Reiners:2010a,Rodler:2011,Figueira:2016} ). As the radial velocity content of a spectrum scales as the resolution to the power $\frac{3}{2}$ \citep{Pepe:2014}, modest errors in stellar line profiles or incomplete line lists may lead to significant errors in the estimation of the RV content of the underlying spectrum. 
For the current analysis,  we used the spectra of Barnard's star in the CRIRES-POP spectral library \citep{Lebzelter:2012}. These observations cover the $Y$, $J$, $H$, $K_s$, $L$ and $M$ bands; of specific interest here being the spectral domain, shortward of 2.38\,$\mu$m, that is amenable to m/s-level precision velocimetry. While the $3-5\mu$m domain contains strong molecular bandpasses that could be of interest for velocimetry, it also suffers from strong telluric absorption and increased thermal background (See \citet{Smette:2015}, Figure~1); due to these challenges and the current lack of effort to develop mid-IR velocimetry, we will not attempt to estimate the relative importance of this wavelength domain. Until recently, this was the only M dwarf with a published near-infrared coverage at very high ($\lambda/\Delta\lambda>70\,000$) resolution.  The recent publication of CARMENES optical and near-infrared spectrum \citep{Reiners:2017} of 324 M dwarfs over the $0.52-1.71\mu$m  domain ($g$ through most of $H$ band) largely fills this gap. The present dataset also includes the $K$ band (1.95-2.40$\mu$m), which is covered by a few pRV spectrograph (SPIRou, GIANO, CRIRES+). None of the currently publicly available high-resolution M dwarf spectrum has been cleaned from telluric absorption through either the near-simultaneous observation of hot stars and/or the combination of exposures taken at varying barycentric velocity.


In an attempt to obtain a realistic estimate of the RV content of M dwarfs and assess the usefulness of existing models, we present a comparison between the spectrum of Barnard's star observed with HARPS, ESPaDOnS and CRIRES \citep{Pepe:2000a, Donati:2006, Kaeufl:2004} and {PHOENIX-ACES} models over the optical and near-infrared domain ($0.4-2.35\mu$m).

We present the properties of Barnard's star in section~\ref{sec:properties} and the dataset used to determine its radial-velocity content in section~\ref{dataset}. Results and discrepancies between the observed and model RV content are presented in section~\ref{section:results} while section~\ref{section:discussion} discusses the implication of these results for nIR PRV instruments.

\section{Barnard's star properties}
\label{sec:properties}

%

\begin{figure}[!th]
\plotone{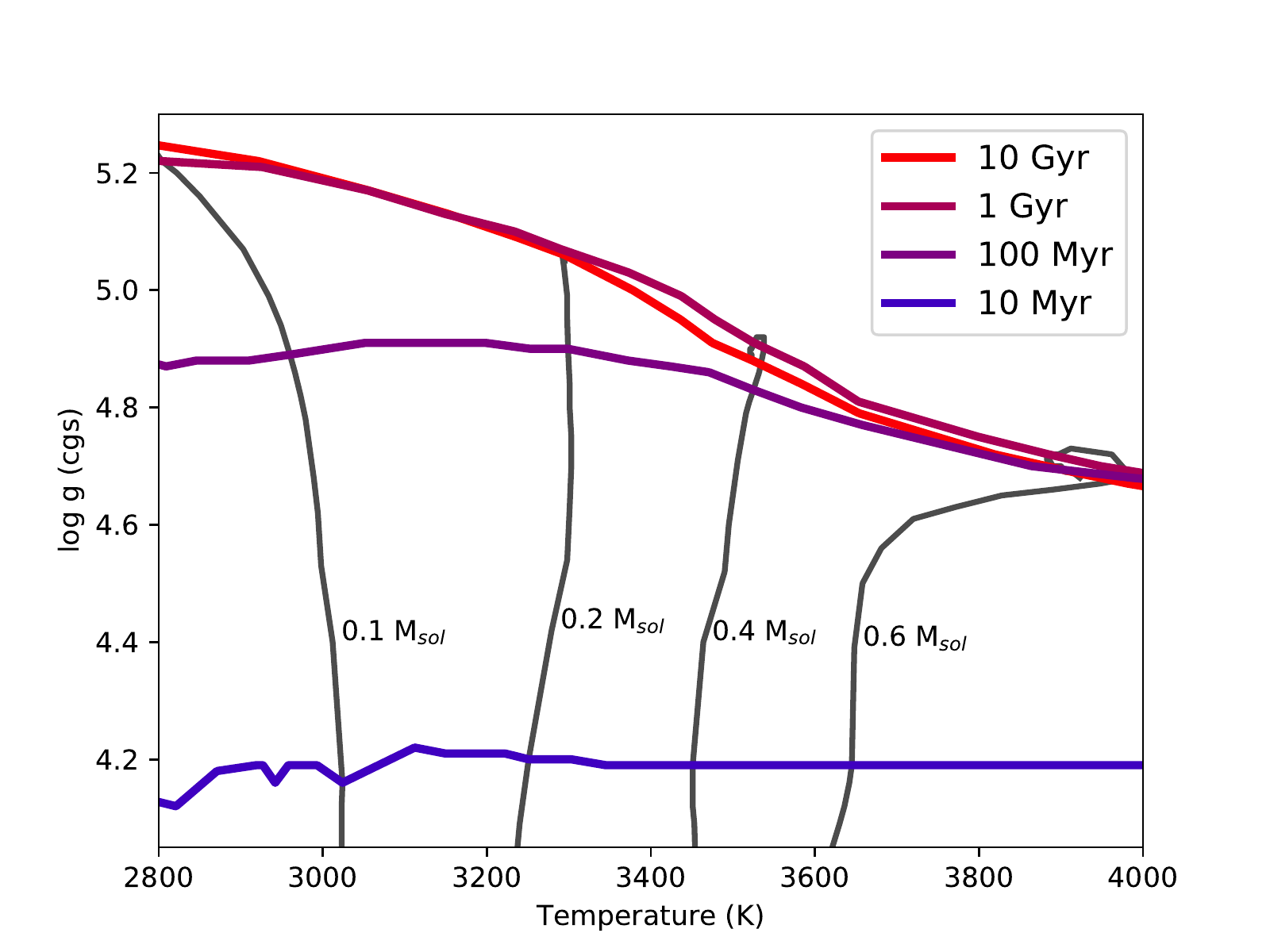}
\caption{\citet{Chabrier:2000} evolutionary models for M dwarfs. At a temperature of 3200\,K and an thick-disk age, Barnard's star is expected to have a $\log g=5.1-5.2$ surface gravity, hence justifying the choice for $\log g=5.0$ atmosphere models. These values are confirmed by observations \citep{Segransan:2003}.}.
 \label{logg:barnard}
 \end{figure}

Barnard's star, at 1.8\,pc from the Sun, is the second closest M dwarf after Proxima Centauri, and it holds the distinction of being the star with the highest apparent proper motion \citep{Barnard:1916}. As a 7-10\,Gyr thick-disk star, it is expected to be a slow rotator; while its rotation period has not been unambiguously established, HST guider photometry points toward a period of $\sim130$\,days \citet{Benedict:1998}. This period is in very good agreement with the estimation of \citealt{Astudillo-Defru:2017} through a determination of $\log{R'_{HK}} = -5.7$. With a $0.2$R$_\odot$ radius, this corresponds to a $v\,\sin i$ smaller than 0.08\,km/s, a negligible contributor compared to the natural line width (thermal, turbulence) or instrumental. { Instrumental broadenings are, at best, on the order of one to a few km/s in the optical (e.g., PEPSI, ESPRESSO respectively with resolutions of up to 1.2 and 2.5\,km/s; \citealt{Strassmeier:2015, Megevand:2014}) or 3-4\,km/s in the near-infrared.} It only shows a modest activity level with occasional flaring activity \citep{Paulson:2006} and its M4V spectral type corresponds to that of the bulk of nearby M dwarfs;  furthermore, Barnard's star spectral type is close to the median of that expected for TESS targets \citep{Sullivan:2015}. Its surface gravity is expected to be slightly above $\log g=5.0$ from evolutionary models (see Figure~\ref{logg:barnard}); this value is overall consistent with measurements of field M surface gravities (e.g., \citealt{Segransan:2003}). Metallicity ([Fe/H]) determination in the litterature range from $-0.13$ to $-0.52$, for the comparison with the model we adopt [Fe/H]$=-0.5$. The properties of Barnard's star are summarized in Table~\ref{prop}. 

As one of our immediate galactic neighbors, this star has been subject to planet searches through astrometry \citep{Benedict:1998}, direct imaging \citep{Gauza:2015a} and radial velocity \citep{Choi:2013, Kurster:2003}, but to date no planet is known around this star and HARPS measurements exclude the existence of planets with a mass superior to $5-6$\,M$_{\oplus}$ in its habitable zone \citep{Bonfils:2013}.

\begin{table}
\begin{center}
\caption{Physical properties of Barnard's star.\label{prop}}
\begin{tabular}{lc}
\hline\hline 
\multicolumn{2}{c}{Other names} \\
\multicolumn{2}{c}{GJ\,699, HIP\,87937, 2MASS\,J17574849+0441405} \\
Spectral type		 & M4\,Ve$^1$\\
Rotation period 	& $\sim$130\,days$^2$ \\
v$\sin i$		& $\le80$\,m/s$^2$ \\

\multicolumn{2}{c}{Radius} \\

\citealt{Segransan:2003}& $0.196\pm0.008$\,R$_\odot$\\
\citealt{Dawson:2004}& $0.200\pm0.008$\,R$_\odot$\\

\multicolumn{2}{c}{Temperature} \\
\citealt{Segransan:2003}& $3163\pm65$\,K \\
\citealt{Dawson:2004}& $3134\pm102$\,K \\
\citealt{Boyajian2012} & $3230\pm10$\,K\\
\citealt{Rojas-Ayala:2012ly} & $3266\pm29$\,K \\ %
\citealt{Neves:2014}& $3338\pm110$\,K \\ 
\citealt{Gaidos:2014}& $3247\pm61$\,K \\ 
\citealt{Mann:2015}& $3228\pm60$\,K \\

Adopted T$_{\rm eff}$ & 3200\,K \\

\multicolumn{2}{c}{Metallicity  [Fe/H]} \\
\citealt{Rojas-Ayala:2012ly} &  $-0.39\pm0.17$\\ 
\citealt{Neves:2013} &$-0.52\pm0.08$\\ 
\citealt{Neves:2014} & $-0.51\pm0.09$\\ 
\citealt{Gaidos:2014} & $-0.32\pm0.08$\\ 
\citealt{Mann:2015} & $-0.40\pm0.08$\\ 
\citealt{Passegger:2016} & $-0.13\pm0.11$\\ 

Adopted metallicity & $-0.5$ \\

\multicolumn{2}{c}{Surface gravity ($\log g$)} \\
\citealt{Segransan:2003} & $5.05\pm0.09$ \\
Adopted $\log g$& $5.0$ \\

\hline\hline
\end{tabular}
\end{center}

$^1$ \citealt{Kirkpatrick:1994}\\
$^2$ \citealt{Benedict:1998}
\end{table}

\section{Datasets and analysis \label{dataset}}

\subsection{The HARPS dataset}
The HARPS high-resolution spectrum used is the median-combination of 22 individual spectrum obtained during a RV planets search \citep{Bonfils:2013}. HARPS \citep{Pepe:2004} is a fiber fed spectrograph at the ESO/3.6-m telescope (La Silla, Chile). It covers the \hbox{380 - 680\,nm} wavelength domain with a resolution of \hbox{$\lambda/\delta\lambda\sim115\,000$}.  We used 104 HARPS spectra from the ESO archive\footnote{IDs 072.C-0488, 183.C-0437} to build a high SNR ($\sim$850 per element) template of Barnard's star. The individual spectra are reprocessed with the latest version of the standard HARPS pipeline \citep{Lovis:2007} which uses nightly set of calibration exposures to locate the orders, flat-field the spectra (Tungsten lamp illumination), and precisely determine the wavelength-calibration scale (ThAr lamp exposure). To build the template we shifted all de-blazed spectra to the rest frame and re-sampled them to a common reference wavelength. We then computed the median flux per spectral element, where the tellurics are discarded from the calculation. As the barycentric Earth radial velocity moves from $-26.3$\,km/s to $26.5$\,km/s in the dataset, the stellar template is free from telluric absorption. This technique is used to produce full templates that are used in cross-correlations for RV measurements in M dwarfs that display rich molecular bands, rather than binary masks that are more appropriate for spectra dominated by atomic lines; see \citealt{Astudillo-Defru:2015} for further details on the construction of this telluric-free template.

\subsection{The ESPaDOnS dataset}

Optical high-resolution spectroscopy of Barnard's star was obtained with ESPaDOnS on 2015, July 30 HST at the Canada-France-Hawaii Telescope (CFHT). Observation was performed using the ``Star \& Sky'' mode combined with the ``normal'' CCD readout mode, to get a resolving power of $R\sim68\,000$ covering the $3700$ to $10\,500$\,\angstrom\, spectral domain over 40 grating orders. The integration time was 90\,s taken at an airmass of 2.2.   The resolution for ESPaDONS is that of the far-red, between the domain covered by HARPS and CRIRES; this value is derived from calibration lamps taken close in time to our observation.

The raw frame was processed by CFHT QSO team using UPENA1.0, an in-house software that calls the {\sc Libre-ESpRIT} pipeline \citealt{Donati:1997}. {\sc Libre-ESpRIT} performs optimal extraction of ESPaDOnS unpolarized (Stokes I) spectrum of the Star and the Sky fibers following the procedure described in \citealt{Donati:1997}. In the present analysis, we used the processed subtracted (Star-Sky) spectrum with a normalized continuum. The accurate wavelength solution that accounts for instrument drifts was measured from strong telluric absorption lines.  ESPaDOnS typically shows drifts well within a resolution element, with typical values below 300\,m/s. At the time of our observations, the measured drift was $-104$\,m/s.

\subsection{CRIRES-POP spectra}

CRIRES spectra were drawn from the CRIRES-POP\footnote{
\href{http://www.univie.ac.at/crirespop/}{http://www.univie.ac.at/crirespop/}} spectral library\citep{Lebzelter:2012}. Individual spectra from the library were analyzed separately and we did not attempt to merge exposure into a single spectrum as we were interested in the shape of line profiles rather than the bulk SED properties. Each spectra was drawn from the library and correlated against a telluric absorption spectrum. Slight offsets in the wavelength calibration (typically $<5$\,km/s) were corrected. We then extracted the time of observation from the file header and determined the barycentric correction for Barnard's star. The CRIRES-POP dataset has recently been used as a test dataset for telluric line subtraction by modeling of absorption \citep{Smette:2015}; while a similar approach could have been applied here to extract RV information from a larger spectral domain, we opted for the simpler approach of performing our analysis on nearly telluric absorption-free  ($<3\%$) parts of the optical and nIR.

\subsection{ G\"ottigen spectral library by Phoenix}

For our analysis, we used PHOENIX-ACES models from the G\"ottigen spectral library\footnote{\href{http://phoenix.astro.physik.uni-goettingen.de}{http://phoenix.astro.physik.uni-goettingen.de}} \citep{Husser:2013a}; these are among the most up-to-date models available and are expected to better represent the nIR spectral features. More specifically, we used the dataset labeled {\it PHOENIX-ACES-AGSS-COND-2011-HiRes}. The model grid is available with a 100\,K temperature step and 0.5\,dex $\log g$ and metallicity steps. For the purpose of comparison with the model, we used a temperature of 3200\,K, a sub-solar metallicity ($-0.5$\,dex) and $\log g=5.0$. Comparison with solar metallicity models ($0.0$\,dex) and low-gravity models ($\log g=4.5$) were also performed in order to assess the impact of varying these parameters on the RV content. The model wavelength grid is finer than the instrumental resolution, which is a necessary condition to properly re-sample models on the wavelength grid of the observations, with a sampling ranging from 0.3 to 0.6\,km/s. This is the same dataset as used by \citet{Figueira:2016}, which leads to a better consistency between the two analysis. 

  In order to account for the finite resolution of instruments, before comparison with observations, models were convolved with the 1-D profile corresponding to that of a circular fiber. The adopted profile corresponds to the profile obtained by collapsing a 2-D circle image onto one axis. For a fiber-fed spectrograph this corresponds to the profile of a monochromatic line in the approximation where the optical design image quality is significantly smaller than the diameter of the fiber. One can show that, arithmetically, this profile corresponds to a $sin$ function between 0 and $\pi$. This profile is representative of most fiber-fed spectrographs (e.g., NIRPS, SPIRou, HARPS). As we are interested in differences between modeled and observed line profiles, we verified that our results were robust against a change in the assume instrumental line-spread-function. In addition to the collapsed-circle profile, We also performed all of the analysis presented here with a gaussian profile having the same FWHM as the collapsed-circle one. All of the conclusions drawn here remain valid with a gaussian profile.


\subsection{Barnard and field M4 photometry\label{subsection:phot}}

In order to derive an RV precision and compare the relative performance reached with various  bandpasses, one needs to scale the flux with photometric measurement. We used the \citet{Mann:2015} $grizJHK$ values for Barnard's Star, but no $Y$-band measurement is available in the literature. We therefore used the  mean $Y-H$ color for M3.5-M5.5 dwarfs in \citet{Hillenbrand:2002} ($Y-H=1.07\pm0.07$, or $Y=5.87\pm0.07$). This allows the scaling of $N_e$ in equation~\ref{eq:RV0}.

 For all comparisons with $z=0$ metallicity models, we use the mean colors for M4V stars in \citealt{Mann:2015}, excluding Barnard's star; see numerical values in Table~\ref{tbl:colors}. These colors are used to scale models and estimate the signal to noise ratio of a given bandpass relative to $J$ band. As expected for a low-metallicity object (e.g., \citealt{Bonfils:2005}), Barnard star has slightly bluer optical to near-infrared colors than field stars; $g-J$ and $r-J$ colors being $\sim0.3$ mag bluer. All colors with $z$, $J$, $H$ and $K_s$ bands are within 0.1\,mag of the field M4V. This is overall consistent with the results from \citealt{Bonfils:2005}, equation 1, where a 0.5\,dex metalicity change corresponds to a 0.27\,mag change in $V-K$. While the strength of molecular bands has a significant impact on the radial velocity content (see Section~\ref{section:qualitative}), the impact of color change is relatively modest, a difference of 0.3\,mag corresponding to a $\sim15$\,\% difference in SNR in the regime where observations are limited by the counting statistics from the source's photons.

\begin{table}[!htbp]
\begin{center}
\caption{Optical and near-infrared colors of M4.0-M4.9 dwarfs in the \citealt{Mann:2015} sample and Barnard Star. The $Y-J$ color is from \citealt{Hillenbrand:2002}, see section~\ref{subsection:phot}. As expected for a low-metallicity object, Barnard's star has slightly bluer optical-to-nIR colors compared to field objects of similar spectral type. \label{tbl:colors}}
\begin{tabular}{|c|cc|}
\hline
color & Field& Barnard \\
\hline
$g-J$ & 5.42  & 5.13 \\
$r-J$ & 3.91  & 3.62 \\
$i-J$ & 2.34  & 2.21 \\
$z-J$ & 1.48  & 1.44 \\
$Y-J$ & 0.50  & 0.50 \\
$J-H$ & 0.56  & 0.49 \\
$H-K_{\rm s}$ & 0.84  & 0.76\\

\hline
\end{tabular}
\end{center}
\end{table}

\subsection{Telluric absorption spectrum}
\label{section:telluric}
Most of the near-infrared domain suffers from absorption by the Earth's atmosphere. Telluric absorption superimposes a set of sharp telluric lines on the stellar spectrum. As the line-of-sight velocity of Barnard's star changes through the year by $\pm32$\,km/s, this component induces a time-varying signal that interferes with precise radial-velocity measurements. Telluric absorption represents a significant challenge to nIR pRV measurements and is discussed at length elsewhere (e.g., \citealt{Artigau:2014c,Bean:2010lr,Seifahrt:2010}). Here, the main problem with telluric absorption in our dataset is that its numerous lines add a significant contribution to the RV content of the spectrum of Barnard's star. We used a model spectrum from the TAPAS\footnote{\href{http://ether.ipsl.jussieu.fr/tapas/}{http://ether.ipsl.jussieu.fr/tapas/}} \citep{Bertaux:2014} for the conditions prevailing at Paranal (airmass of 1, not convolved by an instrumental line width, observation date set as January 1$^{\rm st}$). We included all molecular opacities proposed by the TAPAS interface (Rayleigh, H$_2$O, O$_3$, O$_2$, CO$_2$, CH$_4$ and N$_2$), with an ARLETTY atmospheric model  corresponding to typical conditions occuring in Paranal.  The sampling of the telluric absorption spectrum ranges from 0.2 to 1\,km/s, which is higher than the resolving power of any of our datasets and allows for an accurate interpolation onto the observed wavelength grid. We opted to compare only the radial velocity content of both the observed and model spectra in domains where the atmospheric transmission is $97\%$ or greater. In order to assess the impact of having weak telluric absorption lines contaminating our stellar spectrum, we computed the model RV content with and without multiplying by the TAPAS telluric transmission model. The impact of weak ($<3$\% absorption) telluric lines affects the RV content of the stellar model at the $1$\% level and is deemed negligible in the current analysis.

The exact amount of RV content that can be recovered in the presence of telluric absorption and its impact on the ultimate RV precision is a non-trivial problem (e.g., section~\ref{sec:useful}). Here we are interested in comparing line profile and depth between models and observations, and not the impact of residual telluric absorption on high-precision velocimetry.

\subsection{Useful RV domain in the presence of telluric absorption}
\label{sec:useful}
Masking telluric lines from the stellar spectrum leads to the rejection of part of the wavelength domain that may otherwise be used for radial velocity measurement, provided that efficient subtraction of the telluric absorption contribution can be performed. Various techniques have been proposed to do so: most using atmosphere models to fit telluric absorption (e.g., \citealt{Gullikson:2014,Smette:2015}),  observing reference standard stars of B or A spectral type at roughly the same airmass as the observations \citep{Vacca:2003} or empirical modeling without prior knowledge of telluric absorption \citep{Artigau:2014c}. 

Predicting an RV precision as derived from a model spectrum using a given observational setup in the presence of telluric absorption implies that we assume that telluric absorption will be subtracted up to a certain level. A very conservative approach would reject all of the domain that is affected by telluric absorption at any given time through the year. 
Such drastic wavelength domain rejection is definitely necessary when RV is computed in correlating the stellar spectrum to a reference which is not exactly similar (for example a cross-correlation of the stellar spectra with a numerical weighted mask). However, when the template is similar to the spectra of the star (e.g., median spectrum), only wavelength domain under the telluric lines at the date of the measurement should be rejected. This is well demonstrated in the optical by \citealt{Artigau:2014c} in using RV computation presented in \citealt{Astudillo-Defru:2015} for an early M and a K dwarf.

Whether this holds in the near-infrared remains to be confirmed. As shown in \citet{Artigau:2014c} for $r$-band HARPS observations of an M dwarf, domain with up to 10\% telluric absorption can be used for m/s RV measurements with a proper library of hot star observations. We therefore use this threshold for our RV precision predictions in Section~\ref{section:discussion}, but the aforementioned caveats  apply. To illustrate that our conclusion are only mildly dependent on the exact threshold used for telluric absorption masking, we also computed the RV precision for a much more conservative telluric absorption rejection threshold of $<2\%$. In order to compare the same wavelength domains, models were offset in radial velocity to match to match that of Barnard's star before masking telluric absorption and computing the radial velocity content density ($Q$; see section~\ref{sec:numerical}).



\subsection{  Numerical formalism}
\label{sec:numerical}

In the analysis, we follow the prescription of \citet{Bouchy:2001}. This work evaluates the ultimate precision to which a velocity shift can be determined in a well-sampled spectrum at high signal-to-noise ratio. The RV precision is related to the {\rm quality factor} $Q$ through the relation :

\begin{equation} 
\sigma_{\rm RV} = \frac{c}{Q\sqrt N_e},
\label{eq:RV0}
\end{equation} 

where $c$ is the velocity of light and $N_e$ the number of electrons collected per resolution element, assuming that observations are photon-noise limited (i.e., the effective readout noise per resolution element is much smaller than the photon noise, given by $\sqrt N_e$). As we are interested in comparing the RV precision predicted by models with that of observational data, only the $Q$ value is relevant here as $N_e$ is assumed to be the same. We therefore set :

\begin{equation} 
\sigma_{\rm RV} \propto \frac{1}{Q}.
\label{eq:RV}
\end{equation} 

Following \citealt{Bouchy:2001} notation, $Q$ is 

\begin{equation} 
Q = \frac{\sqrt{\Sigma W(i)}}{\sqrt{\Sigma A_0(i)}} 
\label{eq:sum}
\end{equation} 

with

\begin{equation} 
W(i) = \frac{\lambda^2(i)(\partial A_0(i)/\partial \lambda(i))^2}{A_0(i)};
\label{eq:W}
\end{equation} 

$A_0(i)$ and $\lambda (i)$ respectively denote the flux at a given $(i)$ resolution element and the wavelength of that resolution element. $Q$ is independent of flux, and represents the density of RV content; conversion into an actual RV precision therefore only depends of the total flux ($N_e$). Assuming that the underlying SED is similar within the bandpass of interest, one can therefore directly compare the ratio of $Q$ values to assess the differences in RV content densities between models or observations and models. When comparing the RV precision that one can reach assuming a SNR within a given bandpass, one needs to properly scale the flux (i.e., the $N_e$ term in equation~\ref{eq:RV0}) with actual photometric measurement from the target. As $Q$ is a sum over a given wavelength domain and we are are considering in the spectral distribution of RV content, we will express $Q$ integrated over short wavelength domains. The notation $Q_{\Delta\lambda/\lambda}$ therefore indicates a sum of $Q$ for a running $\Delta\lambda/\lambda$ domain.

\section{Results}
\label{section:results}


\subsection{Barnard's star RV density content}
\label{sec:content}
\begin{figure*}[!htbp]
\includegraphics[width=\linewidth]{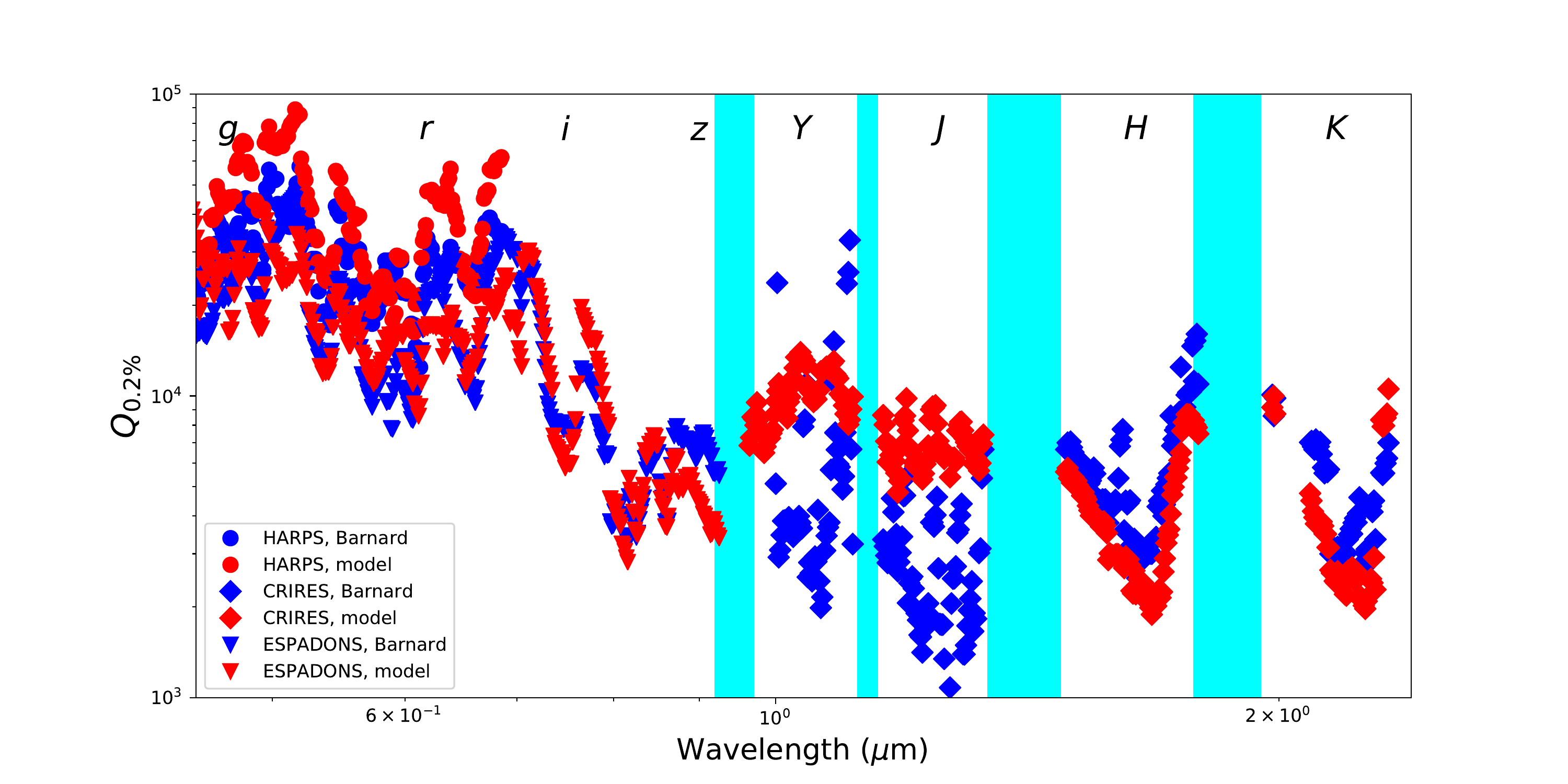}
\includegraphics[width=\linewidth]{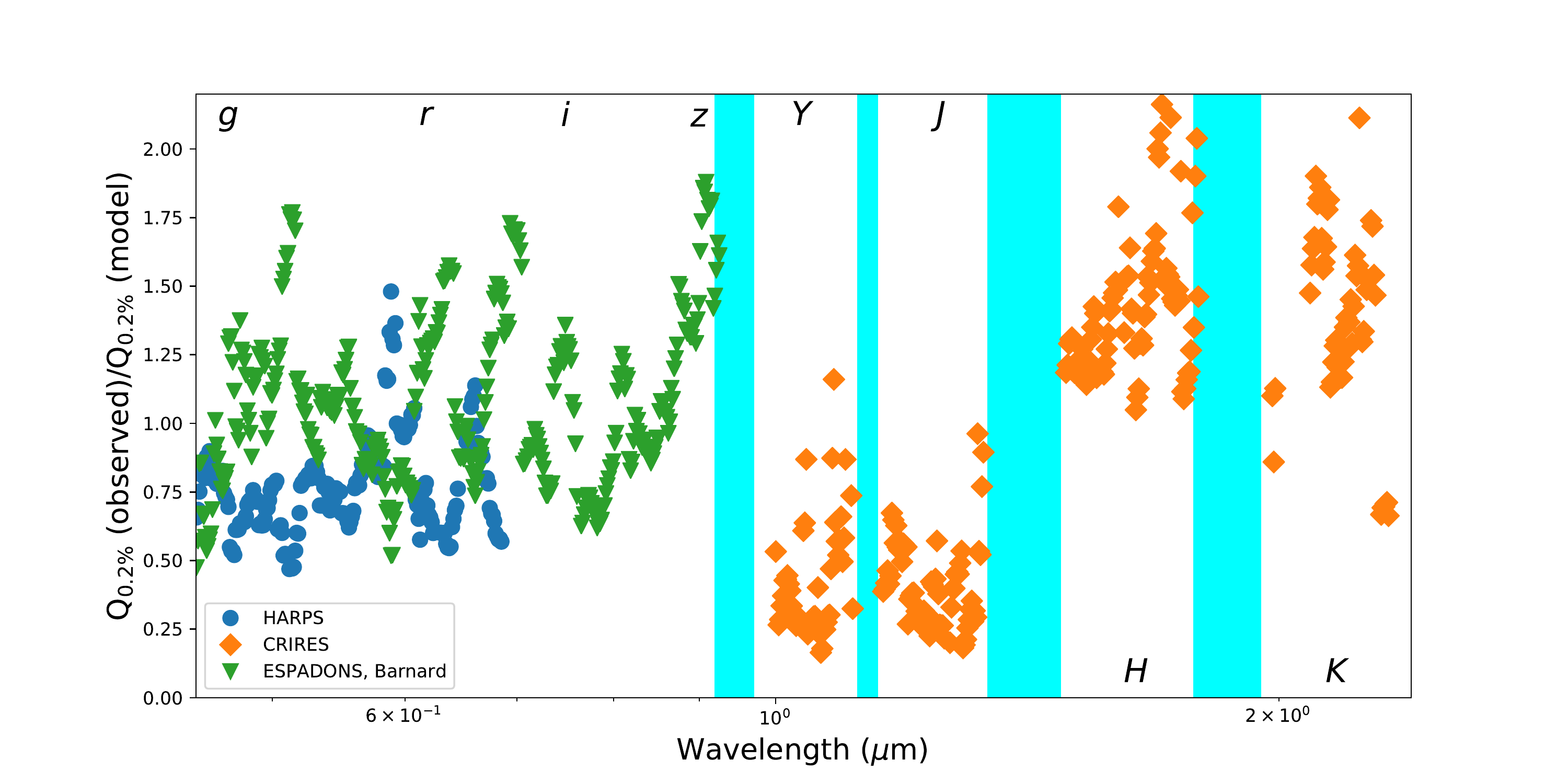}
\caption{(Top) Measured RV content of Barnard's star  over the optical and near-infrared domain. Overall measured (blue) and model (red) RV density are well matched blueward of $\sim1\mu$m. The agreement is poorer in the near-infrared domain with an over-prediction of RV content in $Y$ and $J$ bands and an under-prediction in $H$ and $K$. (Bottom) Ratio of observed to model $Q_{0.2\%}$ values. Areas unusable for RV measurements because of strong telluric absorption are filled in light blue.}
\label{fig:allw}
\end{figure*}

\begin{figure}[!htb]
\includegraphics[width=\linewidth]{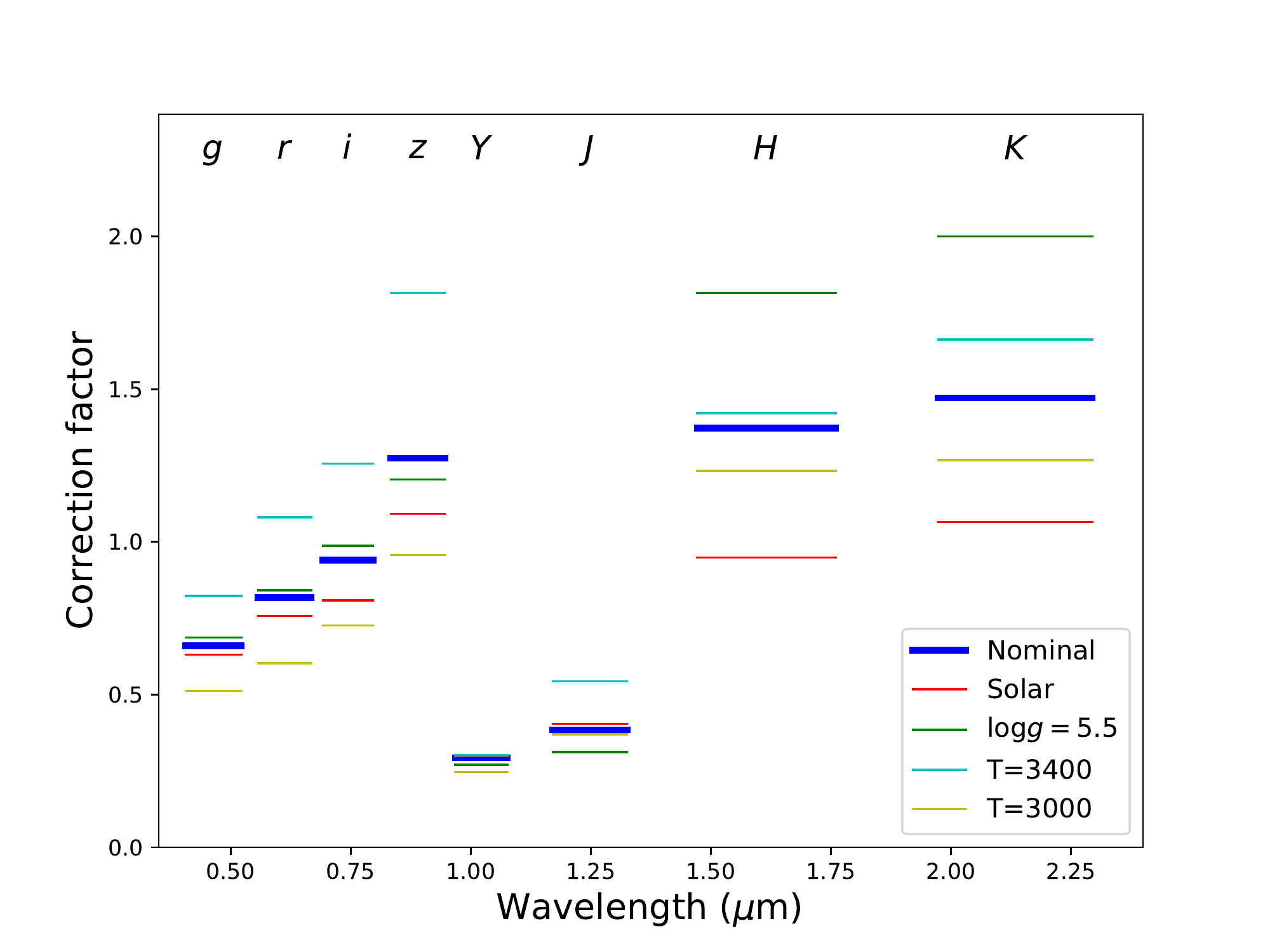}
\caption{Correction factors ($Q_{observed}/Q_{model}$) for the RV precision. The various models tested are described in Section~\ref{section:modelchoice}.}
 \label{fig:correction}
 \end{figure}

The RV content of Barnard's star spectrum was measured from HARPS, ESPaDOnS and CRIRES. The $Q$ value is only computed for telluric-free regions as described in Section~\ref{section:telluric}. The summation as expressed in Equation~\ref{eq:sum} is performed over  $\Delta\lambda/\lambda=0.2\%$ domains through the optical and nIR domain; derived empirical and modeled values are showed in Figure~\ref{fig:allw}. The $Q$ values are globally consistent between models and observations in the optical ($riz$ bandpasses). Near-infrared $Q$ values are much more discrepant, with $Y$ and $J$-band values being over-estimated by models and $H$ and $K$ values under-estimated. From these values, one can determine a median correction to be applied to models to predict the RV precision reachable in the photon-limited regime for all optical to near-infrared bandpasses. The correction corresponds to the flux-weighted mean ratio of $Q_{\rm observed}/Q_{\rm model}$. A correction factor of 0.5 would correspond to an equivalent increase of a factor of 2 in RV error for a given SNR. The precision of the measurement worsens significantly, and this corresponds to a a four-fold loss in observing efficiency (i.e., assuming that signal-to-noise increases as the square root of integration time). Correction values larger than one correspond to an improvement in precision; the RV precision expressed in m/s decreases. Table~\ref{tbl:variants} and Figure~\ref{fig:correction} provides the corresponding relative $Q$ factors $Q_{observed}/Q_{model}$ correction for $grizYJHK$ bandpasses as well as the values derived for different stellar models.

\begin{table}[!htbp]
\begin{center}
\caption{Multiplicative correction factors to be applied on the RV precision derived from stellar models. These values correspond to the square-root of the flux-weighted mean $Q$ ratio between observation and models for each bandpass. The nominal values are for a comparison with the default model described here, but we also explore the impact of other physical parameter choices.\label{tbl:variants}}
\begin{tabular}{|c|c|cccc|}
\hline
 &Nominal&\multicolumn{4}{c|}{Variants}\\
\hline
[Fe/H]&$-0.5$&{\bf 0.0}&$$&$$&$$\\
$\log g$&$5.0$&&{\bf 5.5}&$$&$$\\
$T_{\rm eff}$ (K)&$3200$&&&{\bf 3400}&{\bf 3000}\\
\hline \hline 
$g$ & {\bf   0.66 } &  $ 0.63$ &  $ 0.69$ &  $ 0.82$ &  $ 0.51$\\
$r$ & {\bf   0.82 } &  $ 0.76$ &  $ 0.84$ &  $ 1.08$ &  $ 0.60$\\
$i$ & {\bf   0.94 } &  $ 0.81$ &  $ 0.99$ &  $ 1.26$ &  $ 0.73$\\
$z$ & {\bf   1.27 } &  $ 1.09$ &  $ 1.20$ &  $ 1.82$ &  $ 0.96$\\
$Y$ & {\bf   0.29 } &  $ 0.30$ &  $ 0.27$ &  $ 0.30$ &  $ 0.25$\\
$J$ & {\bf   0.38 } &  $ 0.40$ &  $ 0.31$ &  $ 0.54$ &  $ 0.37$\\
$H$ & {\bf   1.37 } &  $ 0.95$ &  $ 1.82$ &  $ 1.42$ &  $ 1.23$\\
$K$ & {\bf   1.47 } &  $ 1.06$ &  $ 2.00$ &  $ 1.66$ &  $ 1.27$\\

\hline
\end{tabular}
\end{center}
\end{table}

\subsection{Correction value dependence on model choice}
\label{section:modelchoice}
\begin{figure}[!htb]
\plotone{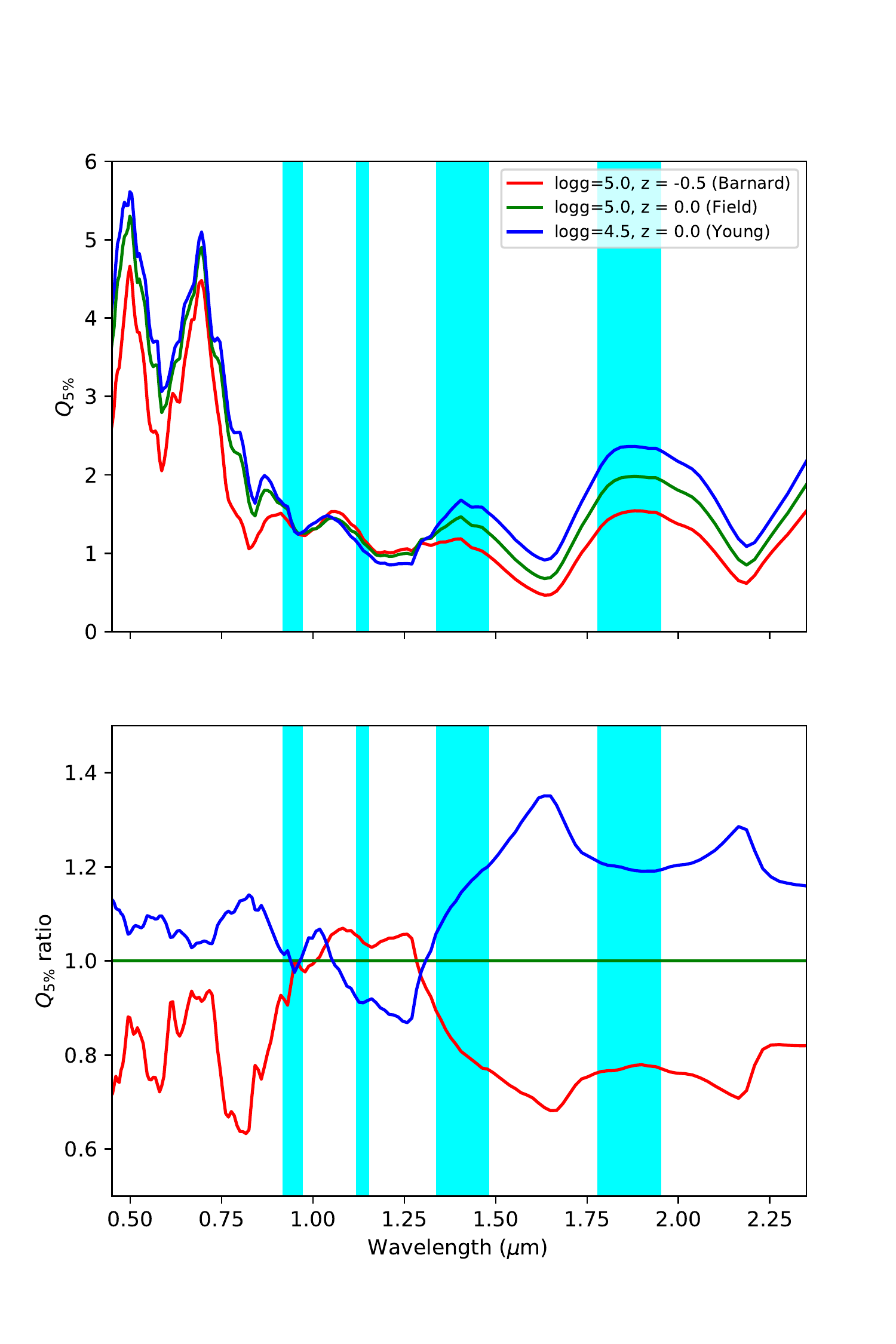}
\caption{ Density of RV content for the optical and near-infrared domain. The top panel shows the RV density $Q_{5\%}$ for 3 models; the nominal Barnard's star model (red), field mid-M at solar metallicity (green) and a low-surface gravity, young, M dwarf (blue). All 3 models are normalized to the $Q_{5\%}$ value of the "Barnard's star" model in $J$. The $i$-band ($\sim0.7\mu$m) contains the highest RV density content, which favors instruments observing in the far-red (See section~\ref{section:discussion}). Solar-metallicity M dwarfs are expected to have a higher RV content than Barnard's star in the near-infrared. The bottom panel shows in red the $Q$ density for the "Barnard's star" and in blue the "Young" models normalized to the "Field" model.}
\label{fig:metal}
\end{figure}

The exact physical parameters of Barnard's star (metallicity, effective temperature, surface gravity) have been measured by several groups and modest discrepancies exist in the literature (See Table~\ref{prop}). It is therefore important to assess whether the results described here hold for different choices of model parameters. The previous results, i.e. that the RV content is over-estimated in $Y$ and $J$ and is under-estimated in $H$ and $K$, remains true if one of the above parameters is changed to one of the extremes of the plausible physical range. Table~\ref{tbl:variants} and Figure~\ref{fig:correction} compile the correction factor that needs to be applied on the RV precision at a given SNR 
for $grizYJHK$ bandpasses as derived from our dataset. The nominal correction applies to a $\log g=5.0$, $-0.5$\,dex metallicity and $T_{\rm eff}=3200$\,K, and corresponds to the nominal model values shown in Figure~\ref{fig:correction}. The values derived when using slightly different models differ, but the overall conclusions remain valid. As shown in Figure~\ref{fig:metal}, models at solar metallicity have a higher $Q$ value in the $H$ and $K$, leading to a more modest correction than for sub-solar metallicity at the same $T_{\rm eff}$. 

When assuming a higher surface gravity ($\log g=5.5$), the correction becomes more important in $H$ and $K$, but has little impact for other bandpasses. Assuming an effective temperature that is hotter or cooler by 200\,K (i.e., a larger difference than suggested by any recent literature value, see Table~\ref{prop}) has little impact on our results. The exact values for the correction are therefore only slightly affected by the choice of physical parameters assumed for Barnard's star and the main conclusions remain unchanged.

\subsection{Qualitative assessment of RV content differences\label{section:qualitative}}

\begin{figure*}[!htbp]
\includegraphics[width=0.52\linewidth]{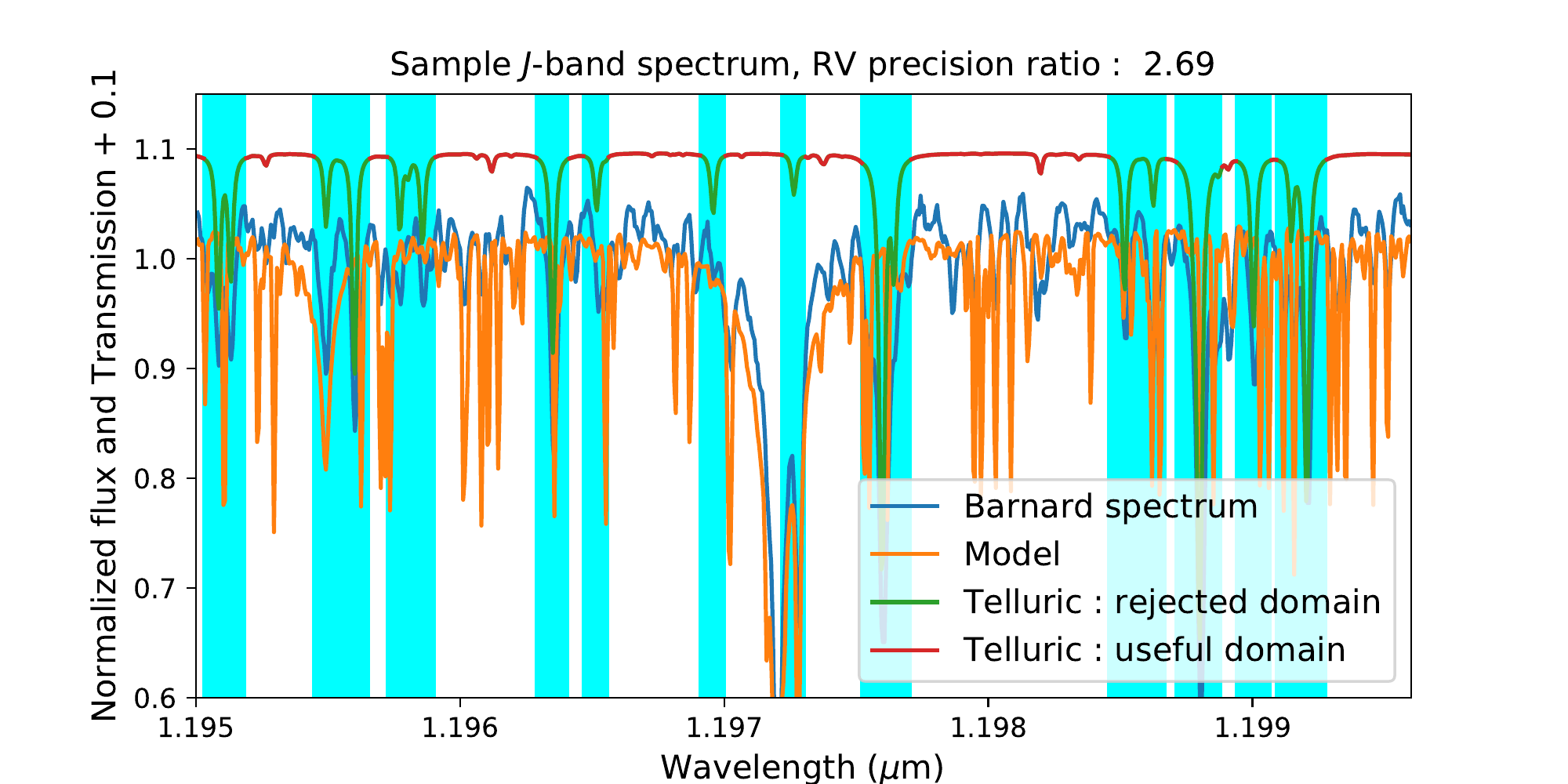}
\includegraphics[width=0.52\linewidth]{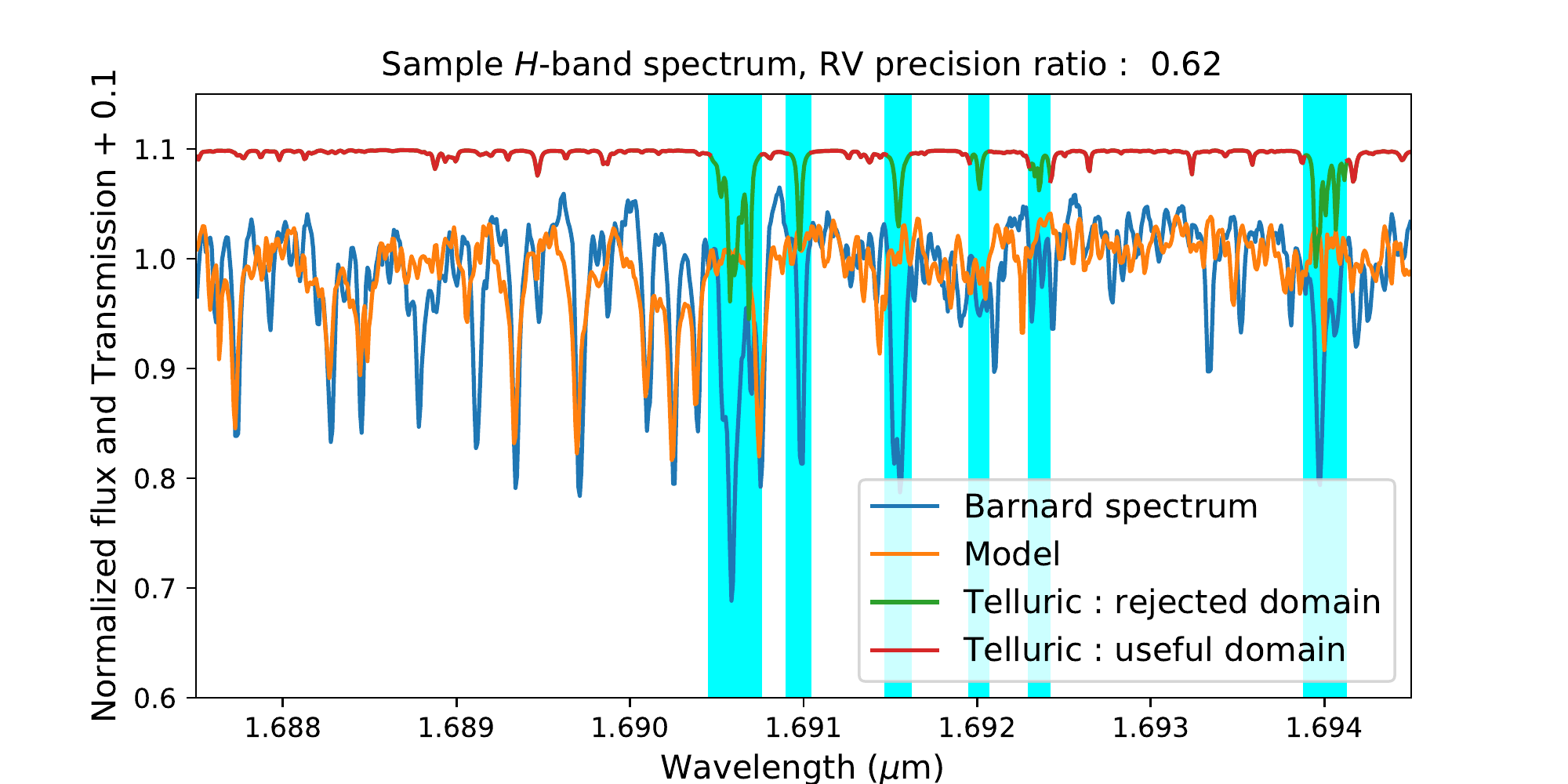}

\caption{Spectrum of Barnard's star from the CRIRES dataset (blue) and models (green) in representative regions of the $J$ (top) and $H$ (bottom) bands. The offset telluric absorption spectrum is shown, with regions included (teal) and excluded (red with cyan background shading) from the determination of $Q$. Within the $J$ band, a large set of lines have over-estimated depth compared to estimations of the RV content. Within $H$, a numerous lines appear to be missing from models, leading to an under-estimation of the RV content.  For the $J$-band sample spectrum domain shown here, the RV precision ratio is $2.69$; for that wavelength domain, the RV precision limit reachable at a given SNR will be degraded by that amount.}
\label{fig:sample}
\end{figure*}

The results we detailed in section~\ref{sec:content} show a significant difference between predicted and observed RV content for Barnard's star in $YJHK$ bands. The difference should lead to notable differences in a direct visual comparison of observed and model spectrum. Figure~\ref{fig:sample} represents two regions of the $J$ and $H$ bands, chosen due to the abundance of sharp lines. The over-estimation in the $J$ band can be traced to deeper and sharper predicted lines than observed. As mentioned earlier, the RV content is proportional to the power $\frac{3}{2}$ of the full-width at half-maximum (FWHM) of lines, so modest differences in line shape leads to significant differences in the predicted RV precision. In $H$ band, numerous lines are observed but not predicted, which is most-likely due to incomplete line lists, as suggested by \cite{Figueira:2016}. It is noteworthy that the RV content is better determined in the optical and far-red, a wavelength domain that has historically received more attention. 

\begin{figure*}[!htbp]
\begin{center}
\includegraphics[width=.245\linewidth]{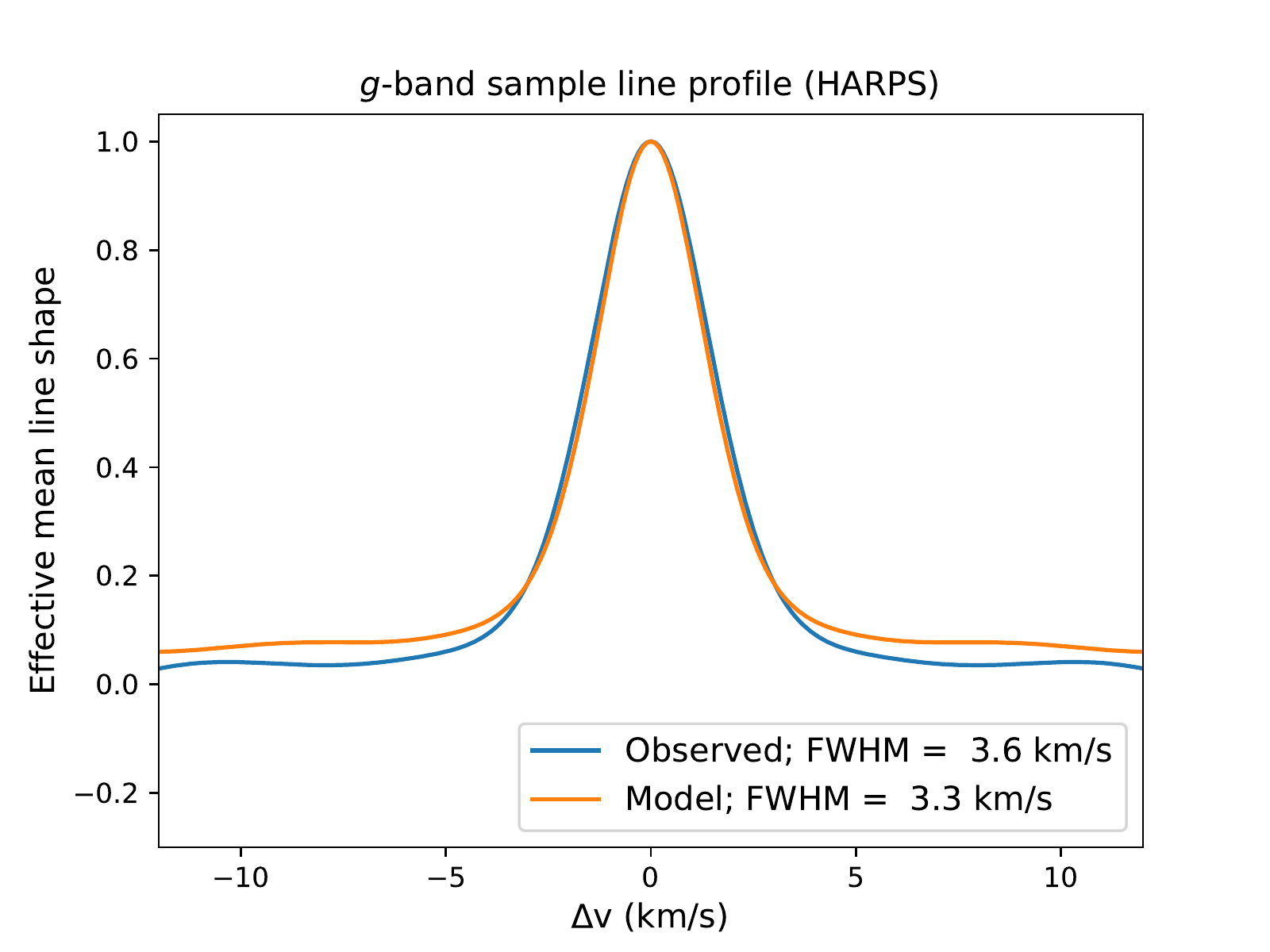}
\includegraphics[width=.245\linewidth]{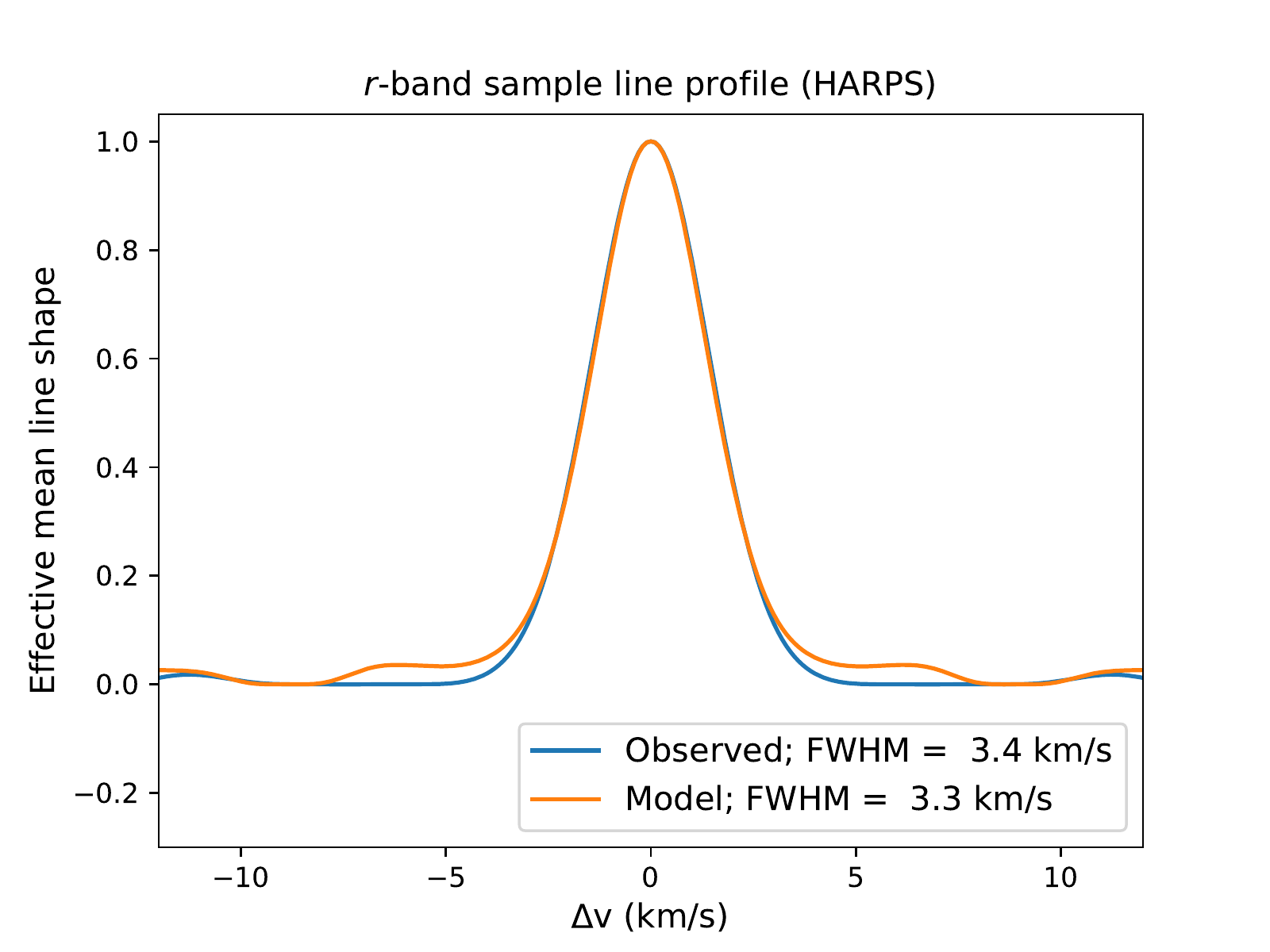}
\includegraphics[width=.245\linewidth]{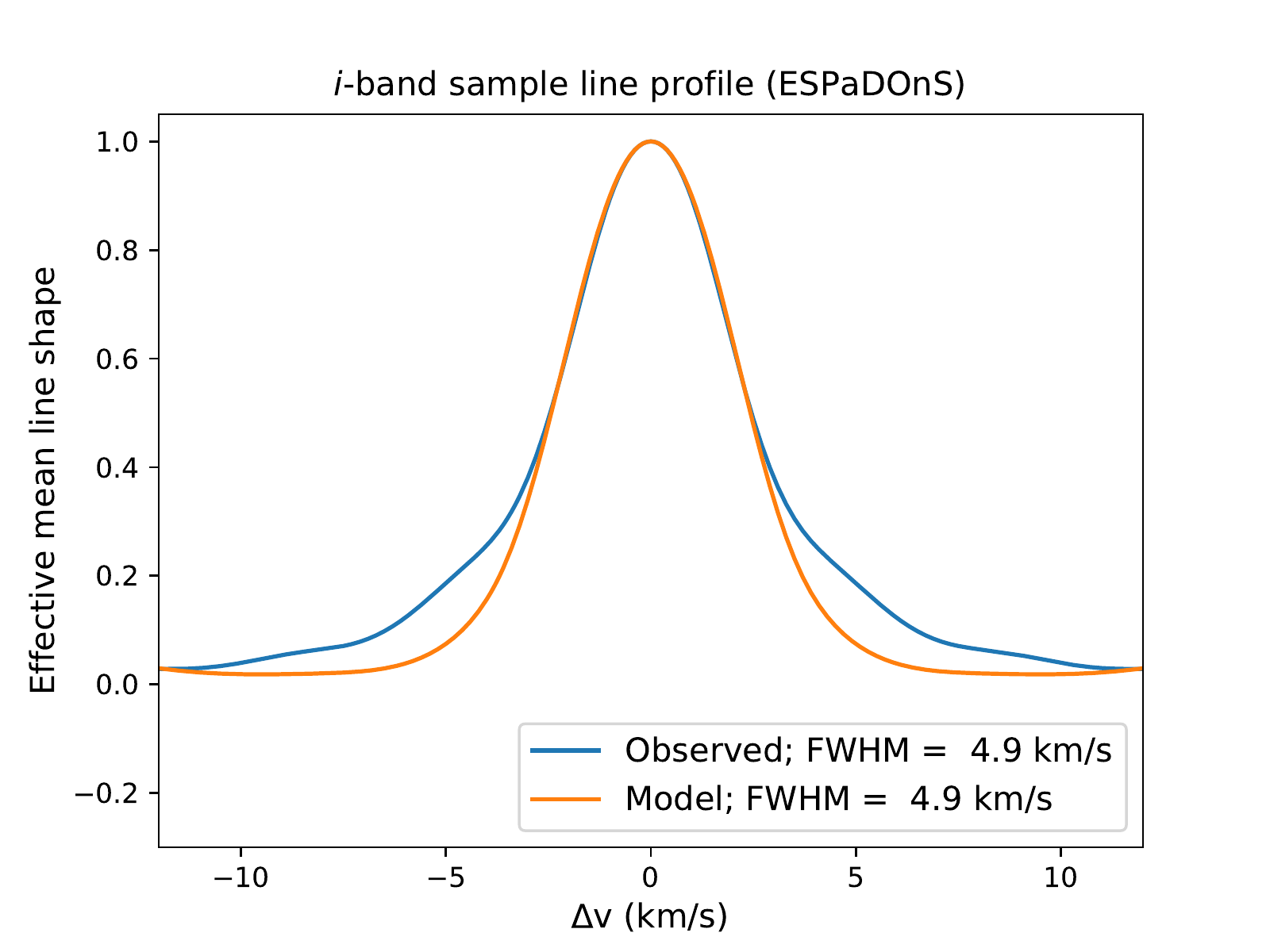}
\includegraphics[width=.245\linewidth]{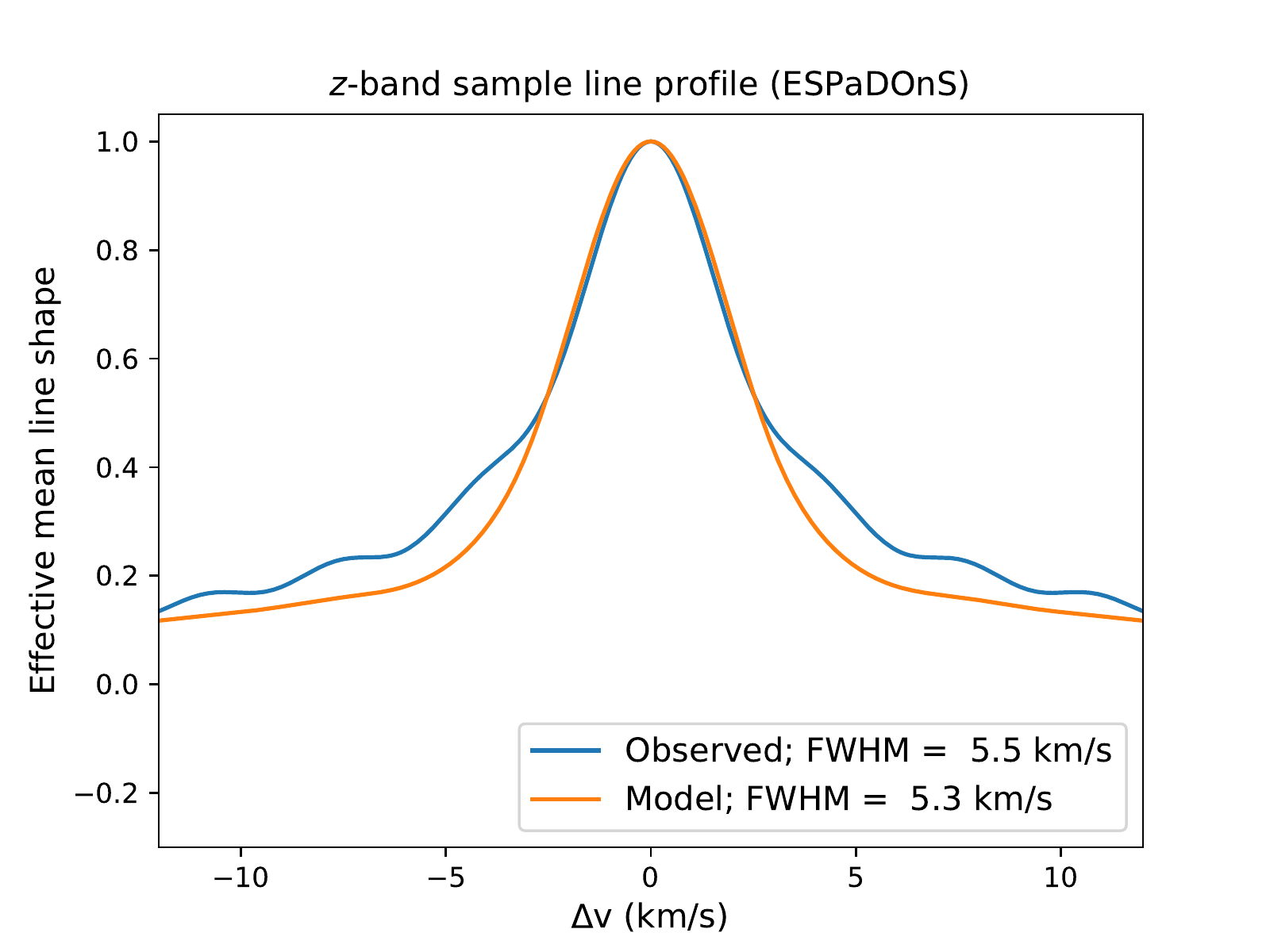}
\includegraphics[width=.245\linewidth]{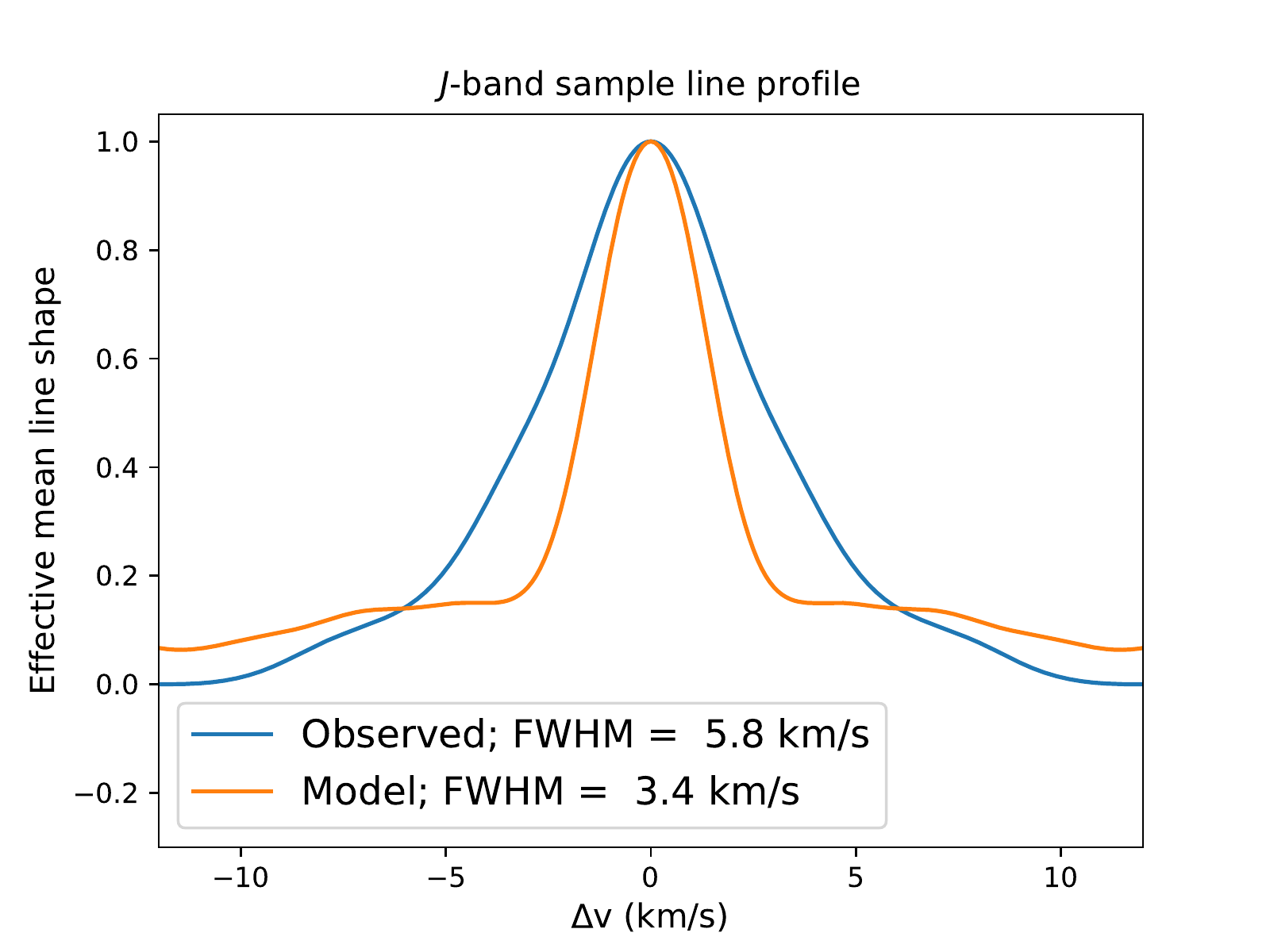}
\includegraphics[width=.245\linewidth]{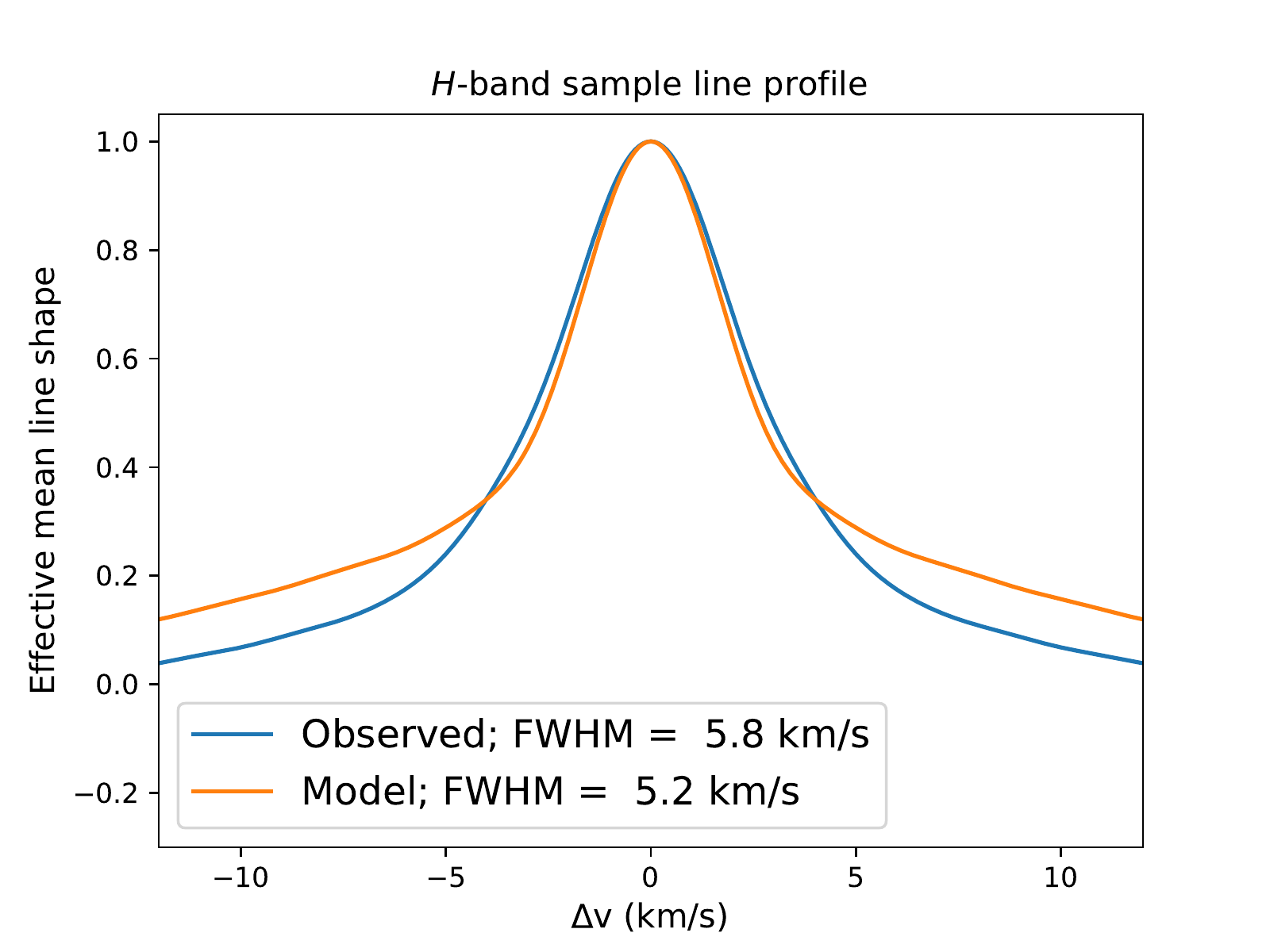}
\includegraphics[width=.245\linewidth]{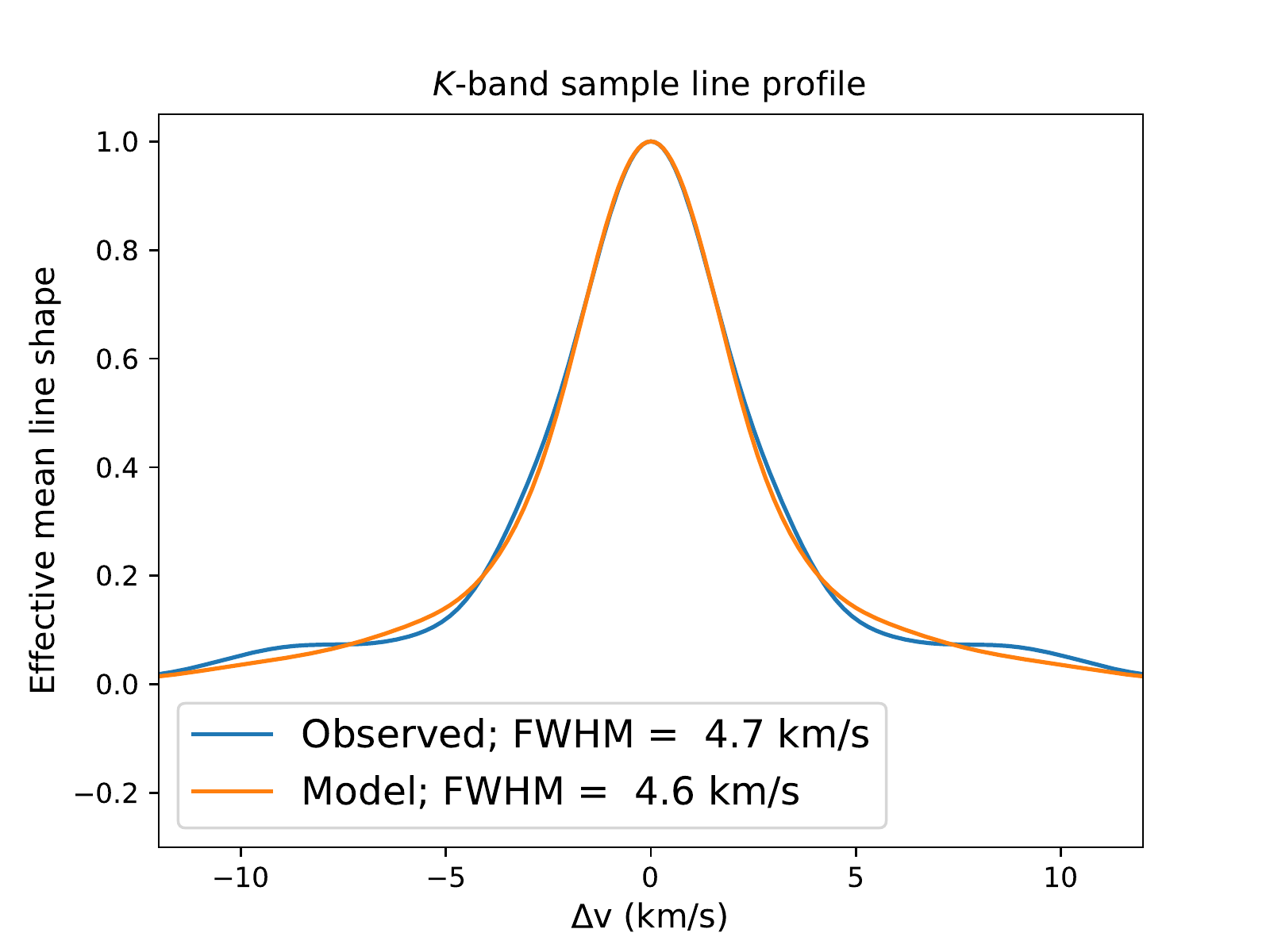}
\end{center}
\caption{Effective mean line profile derived from auto-correlation function of a sample region of $grizJHK$ bands. The profiles are remarkably well matched to the observations, except for $J$ band where much broader profile partially explains the lower RV content compared to models.}
\label{fig:ac}
\end{figure*}

Lines are blended at all wavelengths for M dwarfs and one cannot directly measure an effective line shape directly with isolated lines as can be done for earlier-type stars. We therefore determined the auto-correlation of the spectrum for telluric-free parts of $grizJHK$ bands. The auto-correlation profile of the stellar spectrum is directly linked to the mean line profile, both instrumental and physical. From the auto-correlation profile we recovered the effective mean line profiles (see Figure~\ref{fig:ac}) for both observed and model spectra. In $J$ and $H$, the full-width at half maximum of the line profile is $\sim$5\,km/s, while models predict significantly narrower lines in $J$. This is consistent with the results displayed in Figure~\ref{fig:sample}, where numerous lines are deeper and narrower in models than they are in the observed spectrum, thus leading to an over-estimation of the RV content in $J$.  In the optical domain and $K$ band, the agreement between the observed and model profiles is remarkable. The only notable difference between models and observations are the broader line wings in the $i$ and $z$ bands.

\section{Discussion\label{section:discussion}}

The results presented here allow one to empirically correct RV content predictions from models. The extent of the validity of these correction factors, both in effective temperature and surface gravity, remain to be established with an analysis comparable to the one presented here, but spanning a range of spectral types. If we assume that the $Q_{\rm observed}/Q_{\rm model}$ ratios measured for Barnard's star hold at a solar metallicity, one can predict the RV precision that will be achievable for mid-Ms observed by upcoming nIR RV spectrographs.

\begin{figure}[!htb]
\includegraphics[width=\linewidth]{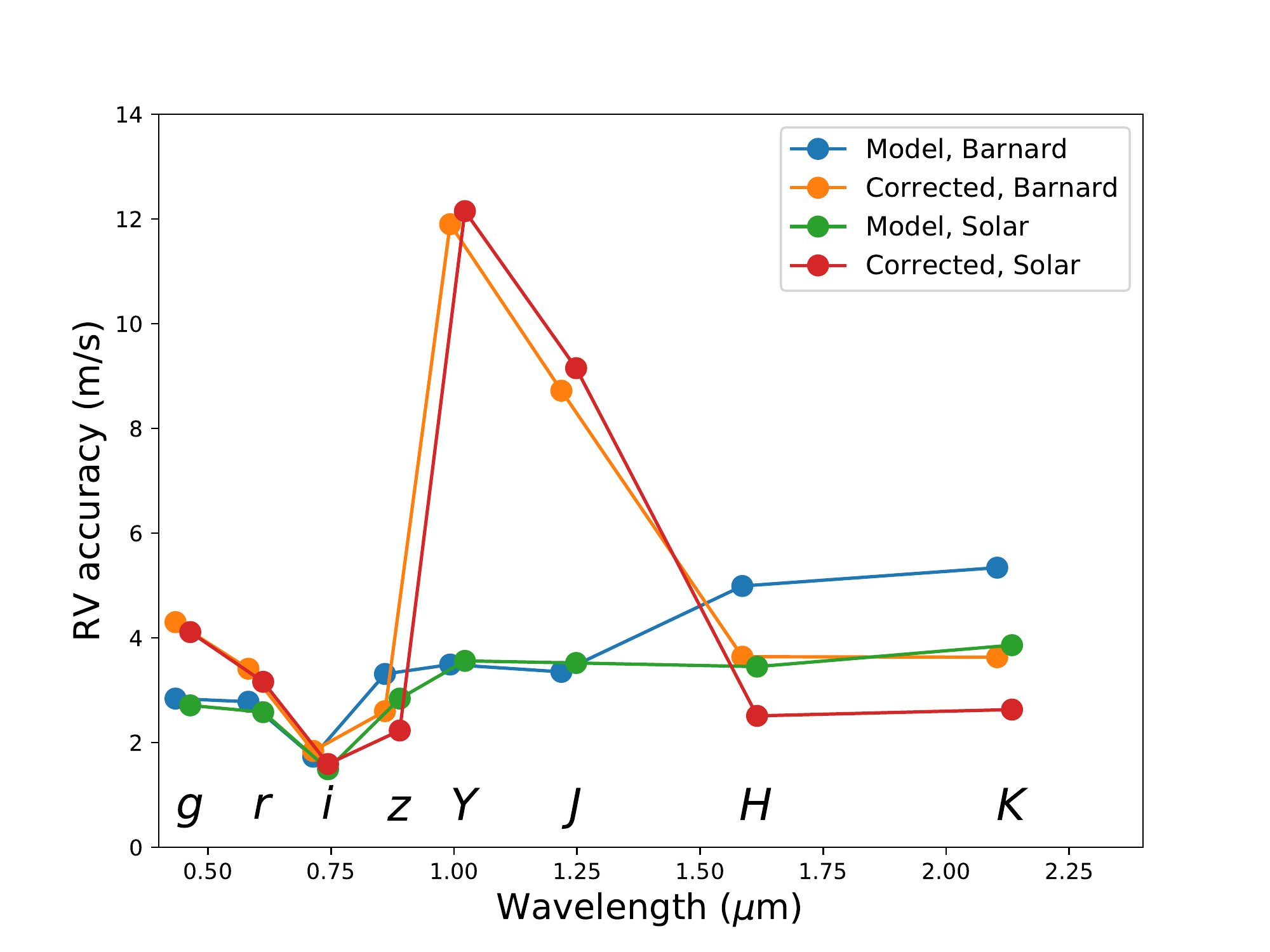}
\caption{ Per-bandpass radial velocity achievable for an SNR=100 in $J$ for a $\lambda/\Delta\lambda=10^5$ resolution element. Models for Barnard's ($\log g=5.0$, metal=$-0.5$\,dex) and Solar ($\log g=5.0$, metal=$0.0$). The correction factor measured for Barnard's star are applied to all models to derive corrected RV precision. Changes are small in the optical between models and corrected values change little the predicted sensitivities. In the near-infrared, models predict comparable contributions from $Y$, $J$, $H$ and $K$ bands for the Solar model (i.e., $\sim3-5$\,m/s), while corrected values indicate much more accurate measurements in $H$ and $K$ compared to $Y$ and $J$.
}
 \label{fig:allrv}
 \end{figure}

We assume that a bandpass contribution to the RV budget scales as $\sigma_{\rm RV}^{-2}$. Two bandpasses that provide a $\sigma_{\rm RV}=1.4$\,m/s contribute as much as a single band for which a $\sigma_{\rm RV}=1$\,m/s measurement is possible in the same amount of time. Figure~\ref{fig:allrv} shows the RV precision per bandpass that is reached for a T$_{\rm eff}=3200$\,K model in 3 metallicity and surface gravity scenarios. The relative contribution of $Y$ and $J$ to the near-infrared RV content budget is predicted to be much smaller than models suggest. For an instrument covering $YJH$ at $R\sim100\,000$ (e.g., NIRPS), predict that $Y$, $J$ and $H$ contribute respectively $39\%$, $42\%$ and $19\%$ of the RV budget. After correction and at solar metallicity, the relative fraction is $7\%$, $14\%$ and $79\%$.

For an instrument covering $YJHK$ domain (e.g., SPIRou, GIANO), the relative contribution of $Y$ and $J$ is even smaller.  Models suggest a similar contributions from all bandpasses ($30\%$, $31\%$, $20\%$, $18\%$) but the correction described here leads to a much larger relative contribution longward of 1.5\,$\mu$m ($3\%$, $6\%$, $45\%$, $47\%$) for solar metallicity. Overall, $H$ and $K$ move from a 38\% to a 94\% contribution to the RV budget.

The importance of $H$ and $K$ band relative to $Y$ and $J$ implies that an RV spectrograph that observes within a single photometric bandpass at a time such as CRIRES+ will be nearly as efficient in $H$ as a similar instrument that would cover the entire $YJH$ domain. The inclusion of $K$ in an instrument such as SPIRou nearly leads to a doubling of the RV content. These results cast a doubt on the conclusion by \citealt{Rodler:2011} that concludes that for M9 and L dwarfs, the most important contribution the the RV content came respectively from $Y$ and $J$. Admittedly our measurement of the RV content of Barnard's star concerns an object $\sim$$1000$\,K hotter, but if the missing opacities in $H$ and $K$ are also present in very-late-Ms and Ls, then these results will need to be revisited. Our results also underline the limitations of works such as \citet{Reiners:2010a} and \citet{Figueira:2016} who, being based on stellar models very similar to the ones presented here, were affected by important systematic errors in the RV estimates.

Having derived correction values for all photometric bandpasses, we can predict the performance for different spectrograph's resolution and $nIR$ domain coverage. We explore the various scenario corresponding to existing and under development PRV spectrographs. Table~\ref{tbl:compare} provides the RV precision reached for a common set of assumptions regarding the target star. As for the above calculation, we assume a mean SNR of 100 per $\Delta\lambda=3\,$km/s  at the center of $J$ band. We did not attempt to provide an exhaustive comparison of the performances of RV spectrographs, an effort that would be much beyond the scope of the current paper. PRV spectrographs are installed on telescope of differing diameter, their overall throughput and intrinsic stability differ and the performances of recently commissioned instruments is likely to improve in the future. Furthermore, depending on the wavelength domain probed, the sensitivity to stellar activity will differ; infrared spectrographs being advantaged, for that matter, relative to optical and far-red PRV instruments (e.g., \citet{Barnes:2011d}). Our comparison therefore only applies to the photon-noise contribution in the complete RV error budget, at a common flux level.

We confirm earlier results (e.g., \citet{Reiners:2010a}, \citet{Seifahrt:2016}) that spectrographs covering the far-red ($griz$ bands) outperform an instrument covering the $YJHK$ domain at the same spectral resolution. In the far-red, the higher RV content density compensates for the lower flux. Interestingly, in such a spectrograph, the $i$ band is more important than $z$ despite the red $i-z$ color of M dwarfs. Qualitatively, this can be seen in Figure~\ref{fig:allw}, where $i$ band has a higher $Q$ value than $z$. \citet{Reiners:2017} presents an analysis of 324 M dwarf spectrum in order to assess their radial-velocity content and its wavelength dependency. There are notable differences between the present analysis of Barnard's star spectrum and that of the representative mid-M shown (e.g., Figure~7 there-in and in particular the M3.5 Luyten’s star). In our analysis, the relative contribution of $J$ and $H$ bands differs significantly while in  \citet{Reiners:2017}, the two bands lead to comparable RV accuracies. Similarly to our results, \citet{Figueira:2016}  predicted a precision much worse for $J$ than for $H$; for the M3 model, $\lambda/\Delta\lambda=80 000$, $v\sin i=1$ and optimal telluric subtraction, the RV accuracy predicted varies from 5\,m/s in $H$ and 16.5\,m/s in $J$ (See Table~A.1 there-in). As pointed in \citet{Reiners:2017}, residual telluric absorption may lead to an increase RV content in their dataset. Residual telluric absorption is also suggested as an explanation for the mismatch between the RV-content based prediction of the RV uncertainties and the measured values.


The lack of M dwarf spectral libraries covering the entire near-infrared at high-resolution until very recently incited previous authors to use models to predict RV content, which, in itself, adds some uncertainties in the interpretation of results. The recent publication of a sample of spectrum obtained with CARMENES \citep{Reiners:2017} partially fills this gap. A need nonetheless remains for a near-infrared spectral atlas cleaned form telluric absorption, either through modeling and/or a combination of multiple observations obtained at sufficiently different barycentric velocities.



\begin{table*}[!htbp]
\begin{center}
\caption{Radial velocity achievable for a T$_{\rm eff}=3200$\,K M dwarf assuming a median SNR=100 in $J$, for a $\lambda/\Delta\lambda=10^5$ resolution element, derived from models with $\log g=5.0$ and $-0.5$\,dex metallicity (Barnard) and, $\log g=5.0$ and 0.0\,dex metallicity (Solar). The corrected values derived from dataset detailed here are given. Values have been computed for the domain within each bandpass with a telluric absorption $<10\%$ (nominal) and $<2\%$ (conservative).  \label{tbl:metal}}
\begin{tabular}{|c|cc|cc|cc|cc|}
\hline
 & 
  \multicolumn{4}{|c|}{ Absorption $<10\%$}
 & 
  \multicolumn{4}{|c|}{ Absorption $<2\%$}
\\
   &\multicolumn{2}{c|}{Barnard}&\multicolumn{2}{c|}{Solar}&\multicolumn{2}{c|}{Barnard}&\multicolumn{2}{c|}{Solar}\\
Band&Model&Corr.&Model&Corr.&Model&Corr.&Model&Corr.\\
& (m/s)&(m/s)&(m/s)&(m/s)& (m/s)&(m/s)&(m/s)&(m/s)\\
\hline
$g$
 &   2.84  &  4.30 &  2.71 &  4.11
 &   2.84  &  4.30 &  2.71 &  4.11
 \\
$r$
 &   2.78  &  3.41 &  2.58 &  3.16
 &   2.79  &  3.41 &  2.58 &  3.16
 \\
$i$
 &   1.73  &  1.84 &  1.49 &  1.59
 &   1.74  &  1.85 &  1.51 &  1.60
 \\
$z$
 &   3.31  &  2.60 &  2.84 &  2.23
 &   3.27  &  2.57 &  2.75 &  2.16
 \\
$Y$
 &   3.49  & 11.90 &  3.56 & 12.15
 &   3.47  & 11.84 &  3.55 & 12.11
 \\
$J$
 &   3.35  &  8.72 &  3.52 &  9.15
 &   3.27  &  8.51 &  3.45 &  8.96
 \\
$H$
 &   4.99  &  3.64 &  3.45 &  2.51
 &   5.14  &  3.75 &  3.54 &  2.58
 \\
$K$
 &   5.34  &  3.63 &  3.86 &  2.63
 &   6.29  &  4.28 &  4.51 &  3.07
 \\

\hline
\end{tabular}
\end{center}
\end{table*}

\begin{table*}[!htbp]
\begin{center}
\caption{ Same as Table~\ref{tbl:metal}, but comparing different instrument resolution and wavelength coverage. The approximate correspondence with existing and planned instrument is also given. These value compare instrumental setups at a common flux level (i.e., SNR=100 in $J$ for an element of \hbox{$\lambda/\Delta\lambda=10^5$})  and only account for difference in the radial velocity content contribution to the error budget. Not taken into account are the intrinsic instrument stability, difference in throughput, sensitivity to stellar activity, etc. Scenarios are designated by the bandpasses they cover and the spectral resolution expressed in thousand (e.g., {\textsc YJH100}) corresponds to a $0.98-1.8\mu$m coverage at R=100\,000.\label{tbl:compare}   }
\begin{tabular}{|c|cc|cc|cc|cc|}
\hline
 & 
  \multicolumn{4}{|c|}{ Absorption $<10\%$}
 & 
  \multicolumn{4}{|c|}{ Absorption $<2\%$}
\\
& 
\multicolumn{2}{c|}{Barnard}&\multicolumn{2}{c|}{Solar} &
\multicolumn{2}{c|}{Barnard}&\multicolumn{2}{c|}{Solar}\\

Scenario &Model&Corr.&Model&Corr.&Model&Corr.&Model&Corr.\\
& (m/s)&(m/s)&(m/s)&(m/s) & (m/s)&(m/s)&(m/s)&(m/s)\\

\hline
{\sc h100}$^{\rm a}$  
  &  4.99 &  3.64  &  3.45 &  2.51
  &  5.14 &  3.75  &  3.54 &  2.58
\\
{\sc k100}$^{\rm b}$  
  &  5.34 &  3.63  &  3.86 &  2.63
  &  6.29 &  4.28  &  4.51 &  3.07
\\
{\sc yjh70}$^{\rm c}$  
  &  3.01 &  4.07  &  2.69 &  2.91
  &  3.00 &  4.16  &  2.70 &  2.98
\\
{\sc yjh80}$^{\rm d}$  
  &  2.60 &  3.65  &  2.36 &  2.64
  &  2.58 &  3.73  &  2.37 &  2.70
\\
{\sc yjh100}$^{\rm e}$  
  &  2.18 &  3.23  &  2.03 &  2.38
  &  2.16 &  3.29  &  2.03 &  2.43
\\
{\sc yjhk50}$^{\rm f}$  
  &  3.35 &  3.48  &  2.81 &  2.51
  &  3.44 &  3.78  &  2.92 &  2.71
\\
{\sc yjhk70}$^{\rm g}$  
  &  2.73 &  2.97  &  2.33 &  2.15
  &  2.79 &  3.23  &  2.42 &  2.32
\\
{\sc riz80}$^{\rm h}$  
  &  1.62 &  1.66  &  1.44 &  1.46
  &  1.62 &  1.66  &  1.44 &  1.46
\\
{\sc gr100}$^{\rm i}$  
  &  1.99 &  2.67  &  1.87 &  2.50
  &  1.99 &  2.67  &  1.87 &  2.50
\\
{\sc gr150}$^{\rm j}$  
  &  1.68 &  2.25  &  1.53 &  2.05
  &  1.68 &  2.25  &  1.53 &  2.05
\\

\hline
\end{tabular}
\end{center}

$^{\rm a,b}$CRIRES+ \citep{Follert:2014},
$^{\rm c}$IRD  \citep{Tamura:2012},
$^{\rm d}$CARMENES-IR \citep{Quirrenbach:2010},
$^{\rm e}$NIRPS\citep{Bouchy:2017},
$^{\rm f}$GIANO \citep{Oliva:2004},
$^{\rm g}$SPIRou \citep{Artigau:2014d},
$^{\rm h}$Maroon-X, CARMENES-optical \citep{Seifahrt:2016, Quirrenbach:2010},
$^{\rm i}$HARPS, HARPS-N \citep{Pepe:2000, Cosentino:2012},
$^{\rm j}$ESPRESSO \citep{Megevand:2014}.
\end{table*}

\acknowledgments

\textit{Acknowledgments}
{\sloppy
The authors would like to thank France Allard for very constructive discussions regarding this work and Xavier Bonfils for their contribution on the optical HARPS spectra of the Barnard's star. CRIRES-POP is based on observations made with ESO telescopes at the La Silla Paranal Observatory under program ID 084.D-0912, 085.D-0161, 086.D-0066, 087.D-0195, and 088.D-0109. Based on observations obtained under program ID 15AC24, at the Canada-France-Hawaii Telescope (CFHT) which is operated by the National Research Council of Canada, the Institut National des Sciences de l'Univers of the Centre National de la Recherche Scientique of France, and the University of Hawaii.
}
{\sloppy
EA and RD acknowledge financial support from the National Science and Engineering Research Council of Canada and the Trottier Family Foundation. PF acknowledges support by Funda\c{c}\~ao para a Ci\^encia e a Tecnologia (FCT) through Investigador FCT contract of reference IF/ 01037/2013, and POPH/FSE (EC) by FEDER funding through the program ``Programa Operacional de Factores de Competitividade - COMPETE''. PF further acknowledges support from FCT in the form of an exploratory project of reference IF/ 01037/2013CP1191/CT0001.   
}
{\sloppy
XD acknowledges the support of the PNP (Programme national de plan\'etologie) and of the Labex OSUG@2020.
}

\bibliography{bibdesk}

\begin{thebibliography}{69}
\expandafter\ifx\csname natexlab\endcsname\relax\def\natexlab#1{#1}\fi

\bibitem[{{Anglada-Escud{\'e}} {et~al.}(2016){Anglada-Escud{\'e}}, {Amado},
  {Barnes}, {Berdi{\~n}as}, {Butler}, {Coleman}, {de La Cueva}, {Dreizler},
  {Endl}, {Giesers}, {Jeffers}, {Jenkins}, {Jones}, {Kiraga}, {K{\"u}rster},
  {L{\'o}pez-Gonz{\'a}lez}, {Marvin}, {Morales}, {Morin}, {Nelson}, {Ortiz},
  {Ofir}, {Paardekooper}, {Reiners}, {Rodr{\'{\i}}guez},
  {Rodr{\'{\i}}guez-L{\'o}pez}, {Sarmiento}, {Strachan}, {Tsapras}, {Tuomi}, \&
  {Zechmeister}}]{Anglada-Escude:2016}
{Anglada-Escud{\'e}}, G., {et~al.} 2016, \nat, 536, 437

\bibitem[{{Artigau} {et~al.}(2014{\natexlab{a}}){Artigau}, {Kouach}, {Donati},
  {Doyon}, {Delfosse}, {Baratchart}, {Lacombe}, {Moutou}, {Rabou}, {Par{\`e}s},
  {Micheau}, {Thibault}, {Reshetov}, {Dubois}, {Hernandez}, {Vall{\'e}e},
  {Wang}, {Dolon}, {Pepe}, {Bouchy}, {Striebig}, {H{\'e}nault}, {Loop},
  {Saddlemyer}, {Barrick}, {Vermeulen}, {Dupieux}, {H{\'e}brard}, {Boisse},
  {Martioli}, {Alencar}, {do Nascimento}, \& {Figueira}}]{Artigau:2014d}
{Artigau}, {\'E}., {et~al.} 2014{\natexlab{a}}, in Society of Photo-Optical
  Instrumentation Engineers (SPIE) Conference Series, Vol. 9147, Society of
  Photo-Optical Instrumentation Engineers (SPIE) Conference Series, 15

\bibitem[{{Artigau} {et~al.}(2014{\natexlab{b}}){Artigau}, {Astudillo-Defru},
  {Delfosse}, {Bouchy}, {Bonfils}, {Lovis}, {Pepe}, {Moutou}, {Donati},
  {Doyon}, \& {Malo}}]{Artigau:2014c}
{Artigau}, {\'E}., {et~al.} 2014{\natexlab{b}}, in Society of Photo-Optical
  Instrumentation Engineers (SPIE) Conference Series, Vol. 9149, Society of
  Photo-Optical Instrumentation Engineers (SPIE) Conference Series, 5

\bibitem[{{Astudillo-Defru} {et~al.}(2017){Astudillo-Defru}, {Delfosse},
  {Bonfils}, {Forveille}, {Lovis}, \& {Rameau}}]{Astudillo-Defru:2017}
{Astudillo-Defru}, N., {Delfosse}, X., {Bonfils}, X., {Forveille}, T., {Lovis},
  C., \& {Rameau}, J. 2017, \aap, 600, A13

\bibitem[{{Astudillo-Defru} {et~al.}(2015){Astudillo-Defru}, {Bonfils},
  {Delfosse}, {S{\'e}gransan}, {Forveille}, {Bouchy}, {Gillon}, {Lovis},
  {Mayor}, {Neves}, {Pepe}, {Perrier}, {Queloz}, {Rojo}, {Santos}, \&
  {Udry}}]{Astudillo-Defru:2015}
{Astudillo-Defru}, N., {et~al.} 2015, \aap, 575, A119

\bibitem[{{Barnard}(1916)}]{Barnard:1916}
{Barnard}, E.~E. 1916, \aj, 29, 181

\bibitem[{{Barnes} {et~al.}(2011){Barnes}, {Jeffers}, \&
  {Jones}}]{Barnes:2011d}
{Barnes}, J.~R., {Jeffers}, S.~V., \& {Jones}, H.~R.~A. 2011, \mnras, 412, 1599

\bibitem[{{Bean} {et~al.}(2010){Bean}, {Seifahrt}, {Hartman}, {Nilsson},
  {Wiedemann}, {Reiners}, {Dreizler}, \& {Henry}}]{Bean:2010lr}
{Bean}, J.~L., {Seifahrt}, A., {Hartman}, H., {Nilsson}, H., {Wiedemann}, G.,
  {Reiners}, A., {Dreizler}, S., \& {Henry}, T.~J. 2010, \apj, 713, 410

\bibitem[{{Benedict} {et~al.}(1998){Benedict}, {McArthur}, {Nelan}, {Story},
  {Whipple}, {Shelus}, {Jefferys}, {Hemenway}, {Franz}, {Wasserman},
  {Duncombe}, {van Altena}, \& {Fredrick}}]{Benedict:1998}
{Benedict}, G.~F., {et~al.} 1998, \aj, 116, 429

\bibitem[{{Bertaux} {et~al.}(2014){Bertaux}, {Lallement}, {Ferron}, {Boonne},
  \& {Bodichon}}]{Bertaux:2014}
{Bertaux}, J.~L., {Lallement}, R., {Ferron}, S., {Boonne}, C., \& {Bodichon},
  R. 2014, \aap, 564, A46

\bibitem[{{Bonfils} {et~al.}(2005){Bonfils}, {Delfosse}, {Udry}, {Santos},
  {Forveille}, \& {S{\'e}gransan}}]{Bonfils:2005}
{Bonfils}, X., {Delfosse}, X., {Udry}, S., {Santos}, N.~C., {Forveille}, T., \&
  {S{\'e}gransan}, D. 2005, \aap, 442, 635

\bibitem[{{Bonfils} {et~al.}(2013){Bonfils}, {Delfosse}, {Udry}, {Forveille},
  {Mayor}, {Perrier}, {Bouchy}, {Gillon}, {Lovis}, {Pepe}, {Queloz}, {Santos},
  {S{\'e}gransan}, \& {Bertaux}}]{Bonfils:2013}
{Bonfils}, X., {et~al.} 2013, \aap, 549, A109

\bibitem[{{Bouchy} {et~al.}(2001){Bouchy}, {Pepe}, \& {Queloz}}]{Bouchy:2001}
{Bouchy}, F., {Pepe}, F., \& {Queloz}, D. 2001, \aap, 374, 733

\bibitem[{{Bouchy} {et~al.}(2017){Bouchy}, {Doyon}, {Artigau}, {Melo},
  {Hernandez}, {Wildi}, {Delfosse}, {Lovis}, {Figueira}, {Canto Martins},
  {Gonz{\'a}lez Hern{\'a}ndez}, {Thibault}, {Reshetov}, {Pepe}, {Santos}, {de
  Medeiros}, {Rebolo}, {Abreu}, {Adibekyan}, {Bandy}, {Benz}, {Blind},
  {Bohlender}, {Boisse}, {Bovay}, {Broeg}, {Brousseau}, {Cabral}, {Chazelas},
  {Cloutier}, {Coelho}, {Conod}, {Cumming}, {Delabre}, {Genolet}, {Hagelberg},
  {Jayawardhana}, {K{\"a}ufl}, {Lafreni{\`e}re}, {de Castro Le{\~a}o}, {Malo},
  {de Medeiros Martins}, {Matthews}, {Metchev}, {Oshagh}, {Ouellet}, {Parro},
  {Rasilla Pi{\~n}eiro}, {Santos}, {Sarajlic}, {Segovia}, {Sordet}, {Udry},
  {Valencia}, {Vall{\'e}e}, {Venn}, {Wade}, \& {Saddlemyer}}]{Bouchy:2017}
{Bouchy}, F., {et~al.} 2017, The Messenger, 169, 21

\bibitem[{{Boyajian} {et~al.}(2012){Boyajian}, {von Braun}, {van Belle},
  {McAlister}, {ten Brummelaar}, {Kane}, {Muirhead}, {Jones}, {White},
  {Schaefer}, {Ciardi}, {Henry}, {L{\'o}pez-Morales}, {Ridgway}, {Gies}, {Jao},
  {Rojas-Ayala}, {Parks}, {Sturmann}, {Sturmann}, {Turner}, {Farrington},
  {Goldfinger}, \& {Berger}}]{Boyajian2012}
{Boyajian}, T.~S., {et~al.} 2012, \apj, 757, 112

\bibitem[{{Chabrier} {et~al.}(2000){Chabrier}, {Baraffe}, {Allard}, \&
  {Hauschildt}}]{Chabrier:2000}
{Chabrier}, G., {Baraffe}, I., {Allard}, F., \& {Hauschildt}, P. 2000, \apj,
  542, 464

\bibitem[{{Chakraborty} {et~al.}(2014){Chakraborty}, {Mahadevan}, {Roy},
  {Dixit}, {Richardson}, {Dongre}, {Pathan}, {Chaturvedi}, {Shah}, {Ubale}, \&
  {Anandarao}}]{Chakraborty:2014}
{Chakraborty}, A., {et~al.} 2014, \pasp, 126, 133

\bibitem[{{Choi} {et~al.}(2013){Choi}, {McCarthy}, {Marcy}, {Howard},
  {Fischer}, {Johnson}, {Isaacson}, \& {Wright}}]{Choi:2013}
{Choi}, J., {McCarthy}, C., {Marcy}, G.~W., {Howard}, A.~W., {Fischer}, D.~A.,
  {Johnson}, J.~A., {Isaacson}, H., \& {Wright}, J.~T. 2013, \apj, 764, 131

\bibitem[{{Cloutier} {et~al.}(2017){Cloutier}, {Astudillo-Defru}, {Doyon},
  {Bonfils}, {Almenara}, {Benneke}, {Bouchy}, {Delfosse}, {Ehrenreich},
  {Forveille}, {Lovis}, {Mayor}, {Menou}, {Murgas}, {Pepe}, {Rowe}, {Santos},
  {Udry}, \& {W{\"u}nsche}}]{Cloutier:2017}
{Cloutier}, R., {et~al.} 2017, ArXiv e-prints

\bibitem[{{Connes}(1985)}]{Connes:1985}
{Connes}, P. 1985, \apss, 110, 211

\bibitem[{{Cosentino} {et~al.}(2012){Cosentino}, {Lovis}, {Pepe}, {Collier
  Cameron}, {Latham}, {Molinari}, {Udry}, {Bezawada}, {Black}, {Born},
  {Buchschacher}, {Charbonneau}, {Figueira}, {Fleury}, {Galli}, {Gallie},
  {Gao}, {Ghedina}, {Gonzalez}, {Gonzalez}, {Guerra}, {Henry}, {Horne},
  {Hughes}, {Kelly}, {Lodi}, {Lunney}, {Maire}, {Mayor}, {Micela}, {Ordway},
  {Peacock}, {Phillips}, {Piotto}, {Pollacco}, {Queloz}, {Rice}, {Riverol},
  {Riverol}, {San Juan}, {Sasselov}, {Segransan}, {Sozzetti}, {Sosnowska},
  {Stobie}, {Szentgyorgyi}, {Vick}, \& {Weber}}]{Cosentino:2012}
{Cosentino}, R., {et~al.} 2012, in \procspie, Vol. 8446, Ground-based and
  Airborne Instrumentation for Astronomy IV, 84461V

\bibitem[{{Dawson} \& {De Robertis}(2004)}]{Dawson:2004}
{Dawson}, P.~C., \& {De Robertis}, M.~M. 2004, \aj, 127, 2909

\bibitem[{{Donati} {et~al.}(2006){Donati}, {Catala}, {Landstreet}, \&
  {Petit}}]{Donati:2006}
{Donati}, J.-F., {Catala}, C., {Landstreet}, J.~D., \& {Petit}, P. 2006, in
  Astronomical Society of the Pacific Conference Series, Vol. 358, Astronomical
  Society of the Pacific Conference Series, ed. R.~{Casini} \& B.~W. {Lites},
  362

\bibitem[{{Donati} {et~al.}(1997){Donati}, {Semel}, {Carter}, {Rees}, \&
  {Collier Cameron}}]{Donati:1997}
{Donati}, J.-F., {Semel}, M., {Carter}, B.~D., {Rees}, D.~E., \& {Collier
  Cameron}, A. 1997, \mnras, 291, 658

\bibitem[{{Dressing} \& {Charbonneau}(2015)}]{Dressing:2015}
{Dressing}, C.~D., \& {Charbonneau}, D. 2015, \apj, 807, 45

\bibitem[{{Figueira} {et~al.}(2016){Figueira}, {Adibekyan}, {Oshagh}, {Neal},
  {Rojas-Ayala}, {Lovis}, {Melo}, {Pepe}, {Santos}, \&
  {Tsantaki}}]{Figueira:2016}
{Figueira}, P., {et~al.} 2016, \aap, 586, A101

\bibitem[{{Fischer} {et~al.}(2016){Fischer}, {Anglada-Escude}, {Arriagada},
  {Baluev}, {Bean}, {Bouchy}, {Buchhave}, {Carroll}, {Chakraborty}, {Crepp},
  {Dawson}, {Diddams}, {Dumusque}, {Eastman}, {Endl}, {Figueira}, {Ford},
  {Foreman-Mackey}, {Fournier}, {Furesz}, {Gaudi}, {Gregory}, {Grundahl},
  {Hatzes}, {Hebrard}, {Herrero}, {Hogg}, {Howard}, {Johnson}, {Jorden},
  {Jurgenson}, {Latham}, {Laughlin}, {Loredo}, {Lovis}, {Mahadevan},
  {McCracken}, {Pepe}, {Perez}, {Phillips}, {Plavchan}, {Prato}, {Quirrenbach},
  {Reiners}, {Robertson}, {Santos}, {Sawyer}, {Segransan}, {Sozzetti},
  {Steinmetz}, {Szentgyorgyi}, {Udry}, {Valenti}, {Wang}, {Wittenmyer}, \&
  {Wright}}]{Fischer:2016}
{Fischer}, D., {et~al.} 2016, ArXiv e-prints

\bibitem[{{Follert} {et~al.}(2014){Follert}, {Dorn}, {Oliva}, {Lizon},
  {Hatzes}, {Piskunov}, {Reiners}, {Seemann}, {Stempels}, {Heiter}, {Marquart},
  {Lockhart}, {Anglada-Escude}, {L{\"o}winger}, {Baade}, {Grunhut}, {Bristow},
  {Klein}, {Jung}, {Ives}, {Kerber}, {Pozna}, {Paufique}, {Kaeufl}, {Origlia},
  {Valenti}, {Gojak}, {Hilker}, {Pasquini}, {Smette}, \&
  {Smoker}}]{Follert:2014}
{Follert}, R., {et~al.} 2014, in \procspie, Vol. 9147, Ground-based and
  Airborne Instrumentation for Astronomy V, 914719

\bibitem[{{Gaidos} {et~al.}(2014){Gaidos}, {Mann}, {L{\'e}pine}, {Buccino},
  {James}, {Ansdell}, {Petrucci}, {Mauas}, \& {Hilton}}]{Gaidos:2014}
{Gaidos}, E., {et~al.} 2014, \mnras, 443, 2561

\bibitem[{{Gauza} {et~al.}(2015){Gauza}, {B{\'e}jar}, {Rebolo}, {{\'A}lvarez},
  {Bihain}, {Zapatero Osorio}, {Caballero}, {Telesco}, \&
  {Packham}}]{Gauza:2015a}
{Gauza}, B., {et~al.} 2015, \mnras, 452, 1677

\bibitem[{{Gullikson} {et~al.}(2014){Gullikson}, {Dodson-Robinson}, \&
  {Kraus}}]{Gullikson:2014}
{Gullikson}, K., {Dodson-Robinson}, S., \& {Kraus}, A. 2014, \aj, 148, 53

\bibitem[{{Hillenbrand} {et~al.}(2002){Hillenbrand}, {Foster}, {Persson}, \&
  {Matthews}}]{Hillenbrand:2002}
{Hillenbrand}, L.~A., {Foster}, J.~B., {Persson}, S.~E., \& {Matthews}, K.
  2002, \pasp, 114, 708

\bibitem[{{Husser} {et~al.}(2013){Husser}, {Wende-von Berg}, {Dreizler},
  {Homeier}, {Reiners}, {Barman}, \& {Hauschildt}}]{Husser:2013a}
{Husser}, T.-O., {Wende-von Berg}, S., {Dreizler}, S., {Homeier}, D.,
  {Reiners}, A., {Barman}, T., \& {Hauschildt}, P.~H. 2013, \aap, 553, A6

\bibitem[{{Kaeufl} {et~al.}(2004){Kaeufl}, {Ballester}, {Biereichel},
  {Delabre}, {Donaldson}, {Dorn}, {Fedrigo}, {Finger}, {Fischer}, {Franza},
  {Gojak}, {Huster}, {Jung}, {Lizon}, {Mehrgan}, {Meyer}, {Moorwood}, {Pirard},
  {Paufique}, {Pozna}, {Siebenmorgen}, {Silber}, {Stegmeier}, \&
  {Wegerer}}]{Kaeufl:2004}
{Kaeufl}, H.-U., {et~al.} 2004, in \procspie, Vol. 5492, Ground-based
  Instrumentation for Astronomy, ed. A.~F.~M. {Moorwood} \& M.~{Iye},
  1218--1227

\bibitem[{{Kirkpatrick} \& {McCarthy}(1994)}]{Kirkpatrick:1994}
{Kirkpatrick}, J.~D., \& {McCarthy}, Jr., D.~W. 1994, \aj, 107, 333

\bibitem[{{K{\"u}rster} {et~al.}(2003){K{\"u}rster}, {Endl}, {Rouesnel}, {Els},
  {Kaufer}, {Brillant}, {Hatzes}, {Saar}, \& {Cochran}}]{Kurster:2003}
{K{\"u}rster}, M., {et~al.} 2003, in ESA Special Publication, Vol. 539, Earths:
  DARWIN/TPF and the Search for Extrasolar Terrestrial Planets, ed.
  M.~{Fridlund}, T.~{Henning}, \& H.~{Lacoste}, 485--489

\bibitem[{{Lebzelter} {et~al.}(2012){Lebzelter}, {Seifahrt}, {Uttenthaler},
  {Ramsay}, {Hartman}, {Nieva}, {Przybilla}, {Smette}, {Wahlgren}, {Wolff},
  {Hussain}, {K{\"a}ufl}, \& {Seemann}}]{Lebzelter:2012}
{Lebzelter}, T., {et~al.} 2012, \aap, 539, A109

\bibitem[{{Lovis} \& {Pepe}(2007)}]{Lovis:2007}
{Lovis}, C., \& {Pepe}, F. 2007, \aap, 468, 1115

\bibitem[{{Lovis} {et~al.}(2016){Lovis}, {Snellen}, {Mouillet}, {Pepe},
  {Wildi}, {Astudillo-Defru}, {Beuzit}, {Bonfils}, {Cheetham}, {Conod},
  {Delfosse}, {Ehrenreich}, {Figueira}, {Forveille}, {Martins}, {Quanz},
  {Santos}, {Schmid}, {S{\'e}gransan}, \& {Udry}}]{Lovis:2016}
{Lovis}, C., {et~al.} 2016, ArXiv e-prints

\bibitem[{{Mahadevan} {et~al.}(2014){Mahadevan}, {Ramsey}, {Terrien},
  {Halverson}, {Roy}, {Hearty}, {Levi}, {Stefansson}, {Robertson}, {Bender},
  {Schwab}, \& {Nelson}}]{Mahadevan:2014}
{Mahadevan}, S., {et~al.} 2014, in \procspie, Vol. 9147, Ground-based and
  Airborne Instrumentation for Astronomy V, 91471G

\bibitem[{{Mann} {et~al.}(2015){Mann}, {Feiden}, {Gaidos}, {Boyajian}, \& {von
  Braun}}]{Mann:2015}
{Mann}, A.~W., {Feiden}, G.~A., {Gaidos}, E., {Boyajian}, T., \& {von Braun},
  K. 2015, \apj, 804, 64

\bibitem[{{Mayor} \& {Queloz}(1995)}]{Mayor:1995}
{Mayor}, M., \& {Queloz}, D. 1995, \nat, 378, 355

\bibitem[{{M{\'e}gevand} {et~al.}(2014){M{\'e}gevand}, {Zerbi}, {Di
  Marcantonio}, {Cabral}, {Riva}, {Abreu}, {Pepe}, {Cristiani}, {Rebolo Lopez},
  {Santos}, {Dekker}, {Aliverti}, {Allende}, {Amate}, {Avila}, {Baldini},
  {Bandy}, {Bristow}, {Broeg}, {Cirami}, {Coelho}, {Conconi}, {Coretti},
  {Cupani}, {D'Odorico}, {De Caprio}, {Delabre}, {Dorn}, {Figueira}, {Fragoso},
  {Galeotta}, {Genolet}, {Gomes}, {Gonz{\'a}lez Hern{\'a}ndez}, {Hughes},
  {Iwert}, {Kerber}, {Landoni}, {Lizon}, {Lovis}, {Maire}, {Mannetta},
  {Martins}, {Molaro}, {Monteiro}, {Moschetti}, {Oliveira}, {Zapatero Osorio},
  {Poretti}, {Rasilla}, {Santana Tschudi}, {Santos}, {Sosnowska}, {Sousa},
  {Tenegi}, {Toso}, {Vanzella}, \& {Viel}}]{Megevand:2014}
{M{\'e}gevand}, D., {et~al.} 2014, in \procspie, Vol. 9147, Ground-based and
  Airborne Instrumentation for Astronomy V, 91471H

\bibitem[{{Neves} {et~al.}(2013){Neves}, {Bonfils}, {Santos}, {Delfosse},
  {Forveille}, {Allard}, \& {Udry}}]{Neves:2013}
{Neves}, V., {Bonfils}, X., {Santos}, N.~C., {Delfosse}, X., {Forveille}, T.,
  {Allard}, F., \& {Udry}, S. 2013, \aap, 551, A36

\bibitem[{{Neves} {et~al.}(2014){Neves}, {Bonfils}, {Santos}, {Delfosse},
  {Forveille}, {Allard}, \& {Udry}}]{Neves:2014}
---. 2014, \aap, 568, A121

\bibitem[{{Oliva} {et~al.}(2004){Oliva}, {Origlia}, {Maiolino}, {Gennari},
  {Biliotti}, {Rossetti}, {Baffa}, {Leone}, {Montegriffo}, {Lolli}, {D'Amato},
  {Bruno}, {Scuderi}, {Ghinassi}, {Gonzalez}, {Lodi}, {Falcini}, {Giani},
  {Marcucci}, \& {Sozzi}}]{Oliva:2004}
{Oliva}, E., {et~al.} 2004, in \procspie, Vol. 5492, Ground-based
  Instrumentation for Astronomy, ed. A.~F.~M. {Moorwood} \& M.~{Iye},
  1274--1279

\bibitem[{{Passegger} {et~al.}(2016){Passegger}, {Wende-von Berg}, \&
  {Reiners}}]{Passegger:2016}
{Passegger}, V.~M., {Wende-von Berg}, S., \& {Reiners}, A. 2016, \aap, 587, A19

\bibitem[{{Paulson} {et~al.}(2006){Paulson}, {Allred}, {Anderson}, {Hawley},
  {Cochran}, \& {Yelda}}]{Paulson:2006}
{Paulson}, D.~B., {Allred}, J.~C., {Anderson}, R.~B., {Hawley}, S.~L.,
  {Cochran}, W.~D., \& {Yelda}, S. 2006, \pasp, 118, 227

\bibitem[{{Pepe} {et~al.}(2014){Pepe}, {Ehrenreich}, \& {Meyer}}]{Pepe:2014}
{Pepe}, F., {Ehrenreich}, D., \& {Meyer}, M.~R. 2014, \nat, 513, 358

\bibitem[{{Pepe} {et~al.}(2000{\natexlab{a}}){Pepe}, {Mayor}, {Benz},
  {Bertaux}, {Sivan}, {Queloz}, \& {Udry}}]{Pepe:2000}
{Pepe}, F., {Mayor}, M., {Benz}, W., {Bertaux}, J.-L., {Sivan}, J.-P.,
  {Queloz}, D., \& {Udry}, S. 2000{\natexlab{a}}, in From Extrasolar Planets to
  Cosmology: The VLT Opening Symposium, ed. J.~{Bergeron} \& A.~{Renzini}, 572

\bibitem[{{Pepe} {et~al.}(2000{\natexlab{b}}){Pepe}, {Mayor}, {Delabre},
  {Kohler}, {Lacroix}, {Queloz}, {Udry}, {Benz}, {Bertaux}, \&
  {Sivan}}]{Pepe:2000a}
{Pepe}, F., {et~al.} 2000{\natexlab{b}}, in \procspie, Vol. 4008, Optical and
  IR Telescope Instrumentation and Detectors, ed. M.~{Iye} \& A.~F. {Moorwood},
  582--592

\bibitem[{{Pepe} {et~al.}(2004){Pepe}, {Mayor}, {Queloz}, {Benz}, {Bonfils},
  {Bouchy}, {Lo Curto}, {Lovis}, {M{\'e}gevand}, {Moutou}, {Naef}, {Rupprecht},
  {Santos}, {Sivan}, {Sosnowska}, \& {Udry}}]{Pepe:2004}
{Pepe}, F., {et~al.} 2004, \aap, 423, 385

\bibitem[{{Quirrenbach} {et~al.}(2010){Quirrenbach}, {Amado}, {Mandel},
  {Caballero}, {Mundt}, {Ribas}, {Reiners}, {Abril}, {Aceituno}, {Afonso},
  {Barrado y Navascues}, {Bean}, {B{\'e}jar}, {Becerril}, {B{\"o}hm},
  {C{\'a}rdenas}, {Claret}, {Colom{\'e}}, {Costillo}, {Dreizler},
  {Fern{\'a}ndez}, {Francisco}, {Galad{\'{\i}}}, {Garrido}, {Gonz{\'a}lez
  Hern{\'a}ndez}, {Gu{\`a}rdia}, {Guenther}, {Guti{\'e}rrez-Soto}, {Joergens},
  {Hatzes}, {Helmling}, {Henning}, {Herrero}, {K{\"u}rster}, {Laun}, {Lenzen},
  {Mall}, {Martin}, {Mart{\'{\i}}n-Ruiz}, {Mirabet}, {Montes}, {Morales},
  {Morales Mu{\~n}oz}, {Moya}, {Naranjo}, {Rabaza}, {Ram{\'o}n}, {Rebolo},
  {Reffert}, {Rodler}, {Rodr{\'{\i}}guez}, {Rodr{\'{\i}}guez Trinidad},
  {Rohloff}, {S{\'a}nchez Carrasco}, {Schmidt}, {Seifert}, {Setiawan},
  {Solano}, {Stahl}, {Storz}, {Su{\'a}rez}, {Thiele}, {Wagner}, {Wiedemann},
  {Zapatero Osorio}, {del Burgo}, {S{\'a}nchez-Blanco}, \&
  {Xu}}]{Quirrenbach:2010}
{Quirrenbach}, A., {et~al.} 2010, in Society of Photo-Optical Instrumentation
  Engineers (SPIE) Conference Series, Vol. 7735, Society of Photo-Optical
  Instrumentation Engineers (SPIE) Conference Series, 13

\bibitem[{{Rajpurohit} {et~al.}(2013){Rajpurohit}, {Reyl{\'e}}, {Allard},
  {Homeier}, {Schultheis}, {Bessell}, \& {Robin}}]{Rajpurohit:2013}
{Rajpurohit}, A.~S., {Reyl{\'e}}, C., {Allard}, F., {Homeier}, D.,
  {Schultheis}, M., {Bessell}, M.~S., \& {Robin}, A.~C. 2013, \aap, 556, A15

\bibitem[{{Reiners} {et~al.}(2010){Reiners}, {Bean}, {Huber}, {Dreizler},
  {Seifahrt}, \& {Czesla}}]{Reiners:2010a}
{Reiners}, A., {Bean}, J.~L., {Huber}, K.~F., {Dreizler}, S., {Seifahrt}, A.,
  \& {Czesla}, S. 2010, \apj, 710, 432

\bibitem[{{Reiners} {et~al.}(2017){Reiners}, {Zechmeister}, {Caballero},
  {Ribas}, {Morales}, {Jeffers}, {Sch{\"o}fer}, {Tal-Or}, {Quirrenbach},
  {Amado}, {Kaminski}, {Seifert}, {Abril}, {Aceituno}, {Alonso-Floriano},
  {Ammler-von Eiff}, {Antona}, {Anglada-Escud{\'e}}, {Anwand-Heerwart},
  {Arroyo-Torres}, {Azzaro}, {Baroch}, {Barrado}, {Bauer}, {Becerril},
  {B{\'e}jar}, {Ben{\'{\i}}tez}, {Berdi{\~n}as}, {Bergond}, {Bl{\"u}mcke},
  {Brinkm{\"o}ller}, {del Burgo}, {Cano}, {C{\'a}rdenas V{\'a}zquez}, {Casal},
  {Cifuentes}, {Claret}, {Colom{\'e}}, {Cort{\'e}s-Contreras}, {Czesla},
  {D{\'{\i}}ez-Alonso}, {Dreizler}, {Feiz}, {Fern{\'a}ndez}, {Ferro},
  {Fuhrmeister}, {Galad{\'{\i}}-Enr{\'{\i}}quez}, {Garcia-Piquer},
  {Garc{\'{\i}}a Vargas}, {Gesa}, {G{\'o}mez}, {Galera}, {Gonz{\'a}lez
  Hern{\'a}ndez}, {Gonz{\'a}lez-Peinado}, {Gr{\"o}zinger}, {Grohnert},
  {Gu{\`a}rdia}, {Guenther}, {Guijarro}, {de Guindos}, {Guti{\'e}rrez-Soto},
  {Hagen}, {Hatzes}, {Hauschildt}, {Hedrosa}, {Helmling}, {Henning}, {Hermelo},
  {Hern{\'a}ndez Arab{\'{\i}}}, {Hern{\'a}ndez Casta{\~n}o}, {Hern{\'a}ndez
  Hernando}, {Herrero}, {Huber}, {Huke}, {Johnson}, {de Juan}, {Kim}, {Klein},
  {Kl{\"u}ter}, {Klutsch}, {K{\"u}rster}, {Lafarga}, {Lamert}, {Lamp{\'o}n},
  {Lara}, {Laun}, {Lemke}, {Lenzen}, {Launhardt}, {L{\'o}pez del Fresno},
  {L{\'o}pez-Gonz{\'a}lez}, {L{\'o}pez-Puertas}, {L{\'o}pez Salas},
  {L{\'o}pez-Santiago}, {Luque}, {Mag{\'a}n Madinabeitia}, {Mall}, {Mancini},
  {Mandel}, {Marfil}, {Mar{\'{\i}}n Molina}, {Maroto}, {Fern{\'a}ndez},
  {Mart{\'{\i}}n}, {Mart{\'{\i}}n-Ruiz}, {Marvin}, {Mathar}, {Mirabet},
  {Montes}, {Moreno-Raya}, {Moya}, {Mundt}, {Nagel}, {Naranjo}, {Nortmann},
  {Nowak}, {Ofir}, {Oreiro}, {Pall{\'e}}, {Panduro}, {Pascual}, {Passegger},
  {Pavlov}, {Pedraz}, {P{\'e}rez-Calpena}, {P{\'e}rez Medialdea}, {Perger},
  {Perryman}, {Pluto}, {Rabaza}, {Ram{\'o}n}, {Rebolo}, {Redondo}, {Reffert},
  {Reinhart}, {Rhode}, {Rix}, {Rodler}, {Rodr{\'{\i}}guez},
  {Rodr{\'{\i}}guez-L{\'o}pez}, {Rodr{\'{\i}}guez Trinidad}, {Rohloff},
  {Rosich}, {Sadegi}, {S{\'a}nchez-Blanco}, {S{\'a}nchez Carrasco},
  {S{\'a}nchez-L{\'o}pez}, {Sanz-Forcada}, {Sarkis}, {Sarmiento},
  {Sch{\"a}fer}, {Schmitt}, {Schiller}, {Schweitzer}, {Solano}, {Stahl},
  {Strachan}, {St{\"u}rmer}, {Su{\'a}rez}, {Tabernero}, {Tala}, {Trifonov},
  {Tulloch}, {Ulbrich}, {Veredas}, {Vico Linares}, {Vilardell}, {Wagner},
  {Winkler}, {Wolthoff}, {Xu}, {Yan}, \& {Zapatero Osorio}}]{Reiners:2017}
{Reiners}, A., {et~al.} 2017, ArXiv e-prints

\bibitem[{{Ricker} {et~al.}(2014){Ricker}, {Winn}, {Vanderspek}, {Latham},
  {Bakos}, {Bean}, {Berta-Thompson}, {Brown}, {Buchhave}, {Butler}, {Butler},
  {Chaplin}, {Charbonneau}, {Christensen-Dalsgaard}, {Clampin}, {Deming},
  {Doty}, {De Lee}, {Dressing}, {Dunham}, {Endl}, {Fressin}, {Ge}, {Henning},
  {Holman}, {Howard}, {Ida}, {Jenkins}, {Jernigan}, {Johnson}, {Kaltenegger},
  {Kawai}, {Kjeldsen}, {Laughlin}, {Levine}, {Lin}, {Lissauer}, {MacQueen},
  {Marcy}, {McCullough}, {Morton}, {Narita}, {Paegert}, {Palle}, {Pepe},
  {Pepper}, {Quirrenbach}, {Rinehart}, {Sasselov}, {Sato}, {Seager},
  {Sozzetti}, {Stassun}, {Sullivan}, {Szentgyorgyi}, {Torres}, {Udry}, \&
  {Villasenor}}]{Ricker:2014}
{Ricker}, G.~R., {et~al.} 2014, ArXiv e-prints

\bibitem[{{Rodler} {et~al.}(2011){Rodler}, {Del Burgo}, {Witte}, {Helling},
  {Hauschildt}, {Mart{\'{\i}}n}, {{\'A}lvarez}, \& {Deshpande}}]{Rodler:2011}
{Rodler}, F., {Del Burgo}, C., {Witte}, S., {Helling}, C., {Hauschildt}, P.~H.,
  {Mart{\'{\i}}n}, E.~L., {{\'A}lvarez}, C., \& {Deshpande}, R. 2011, \aap,
  532, A31

\bibitem[{{Rojas-Ayala} {et~al.}(2012){Rojas-Ayala}, {Covey}, {Muirhead}, \&
  {Lloyd}}]{Rojas-Ayala:2012ly}
{Rojas-Ayala}, B., {Covey}, K.~R., {Muirhead}, P.~S., \& {Lloyd}, J.~P. 2012,
  \apj, 748, 93

\bibitem[{{S{\'e}gransan} {et~al.}(2003){S{\'e}gransan}, {Kervella},
  {Forveille}, \& {Queloz}}]{Segransan:2003}
{S{\'e}gransan}, D., {Kervella}, P., {Forveille}, T., \& {Queloz}, D. 2003,
  \aap, 397, L5

\bibitem[{{Seifahrt} {et~al.}(2016){Seifahrt}, {Bean}, {St{\"u}rmer}, {Gers},
  {Grobler}, {Reed}, \& {Jones}}]{Seifahrt:2016}
{Seifahrt}, A., {Bean}, J.~L., {St{\"u}rmer}, J., {Gers}, L., {Grobler}, D.~S.,
  {Reed}, T., \& {Jones}, D.~J. 2016, in \procspie, Vol. 9908, Society of
  Photo-Optical Instrumentation Engineers (SPIE) Conference Series, 990818

\bibitem[{{Seifahrt} {et~al.}(2010){Seifahrt}, {K{\"a}ufl}, {Z{\"a}ngl},
  {Bean}, {Richter}, \& {Siebenmorgen}}]{Seifahrt:2010}
{Seifahrt}, A., {K{\"a}ufl}, H.~U., {Z{\"a}ngl}, G., {Bean}, J.~L., {Richter},
  M.~J., \& {Siebenmorgen}, R. 2010, \aap, 524, A11

\bibitem[{{Smette} {et~al.}(2015){Smette}, {Sana}, {Noll}, {Horst}, {Kausch},
  {Kimeswenger}, {Barden}, {Szyszka}, {Jones}, {Gallenne}, {Vinther},
  {Ballester}, \& {Taylor}}]{Smette:2015}
{Smette}, A., {et~al.} 2015, \aap, 576, A77

\bibitem[{{Snellen} {et~al.}(2015){Snellen}, {de Kok}, {Birkby}, {Brandl},
  {Brogi}, {Keller}, {Kenworthy}, {Schwarz}, \& {Stuik}}]{Snellen:2015}
{Snellen}, I., {et~al.} 2015, \aap, 576, A59

\bibitem[{{Strassmeier} {et~al.}(2015){Strassmeier}, {Ilyin}, {J{\"a}rvinen},
  {Weber}, {Woche}, {Barnes}, {Bauer}, {Beckert}, {Bittner}, {Bredthauer},
  {Carroll}, {Denker}, {Dionies}, {DiVarano}, {D{\"o}scher}, {Fechner},
  {Feuerstein}, {Granzer}, {Hahn}, {Harnisch}, {Hofmann}, {Lesser}, {Paschke},
  {Pankratow}, {Plank}, {Pl{\"u}schke}, {Popow}, \&
  {Sablowski}}]{Strassmeier:2015}
{Strassmeier}, K.~G., {et~al.} 2015, Astronomische Nachrichten, 336, 324

\bibitem[{{Sullivan} {et~al.}(2015){Sullivan}, {Winn}, {Berta-Thompson},
  {Charbonneau}, {Deming}, {Dressing}, {Latham}, {Levine}, {McCullough},
  {Morton}, {Ricker}, {Vanderspek}, \& {Woods}}]{Sullivan:2015}
{Sullivan}, P.~W., {et~al.} 2015, \apj, 809, 77

\bibitem[{{Tamura} {et~al.}(2012){Tamura}, {Suto}, {Nishikawa}, {Kotani},
  {Sato}, {Aoki}, {Usuda}, {Kurokawa}, {Kashiwagi}, {Nishiyama}, {Ikeda},
  {Hall}, {Hodapp}, {Hashimoto}, {Morino}, {Inoue}, {Mizuno}, {Washizaki},
  {Tanaka}, {Suzuki}, {Kwon}, {Suenaga}, {Oh}, {Narita}, {Kokubo}, {Hayano},
  {Izumiura}, {Kambe}, {Kudo}, {Kusakabe}, {Ikoma}, {Hori}, {Omiya}, {Genda},
  {Fukui}, {Fujii}, {Guyon}, {Harakawa}, {Hayashi}, {Hidai}, {Hirano},
  {Kuzuhara}, {Machida}, {Matsuo}, {Nagata}, {Ohnuki}, {Ogihara}, {Oshino},
  {Suzuki}, {Takami}, {Takato}, {Takahashi}, {Tachinami}, \&
  {Terada}}]{Tamura:2012}
{Tamura}, M., {et~al.} 2012, in Proc. SPIE, Vol. 8446, Proc. SPIE

\bibitem[{{Vacca} {et~al.}(2003){Vacca}, {Cushing}, \& {Rayner}}]{Vacca:2003}
{Vacca}, W.~D., {Cushing}, M.~C., \& {Rayner}, J.~T. 2003, \pasp, 115, 389

\bibitem[{{Vogt} {et~al.}(1994){Vogt}, {Allen}, {Bigelow}, {Bresee}, {Brown},
  {Cantrall}, {Conrad}, {Couture}, {Delaney}, {Epps}, {Hilyard}, {Hilyard},
  {Horn}, {Jern}, {Kanto}, {Keane}, {Kibrick}, {Lewis}, {Osborne},
  {Pardeilhan}, {Pfister}, {Ricketts}, {Robinson}, {Stover}, {Tucker}, {Ward},
  \& {Wei}}]{Vogt:1994}
{Vogt}, S.~S., {et~al.} 1994, in \procspie, Vol. 2198, Instrumentation in
  Astronomy VIII, ed. D.~L. {Crawford} \& E.~R. {Craine}, 362

\end{thebibliography}

\expandafter\ifx\csname natexlab\endcsname\relax\def\natexlab#1{#1}\fi

\end{document}